%% file: Simulation.tex
\documentclass[11pt]{article}

\usepackage{amssymb,amsmath,amsthm}	
\usepackage{a4wide}
\usepackage[utf8x]{luainputenc}               	
\usepackage{graphicx}      										
\usepackage{tikz}          										
\usepackage[authoryear]{natbib}
\usepackage{dsfont}
\usepackage{hyperref}
\usepackage{multirow}
\usepackage{subcaption}
\usepackage{amsmath}
\usepackage{xspace}
\usepackage{cleveref}
\usepackage{ulem}
\usepackage{natbib}
\usepackage{todonotes}
\usepackage{afterpage}
\usepackage{placeins}

\setcitestyle{round,semicolon,aysep={}}

\newtheorem{definition}{Definition}[]

\newcommand{\R}{\ensuremath{\mathbb{R}}}

\newcommand{\N}{\ensuremath{\mathbb{N}}}

\DeclareMathOperator{\Unifd}{Unif_{d}}
\DeclareMathOperator{\Unifc}{Unif_{c}}

\newcommand{\changes}[1]{\textcolor{black}{#1}}

\usepackage{soul}

\begin{document}
	

	\title{Simulation of Stylized Facts in Agent-Based Computational Economic Market Models }
	
\author{Maximilian Beikirch\footnote{RWTH Aachen University, Templergraben 55, 52056 Aachen, Germany} \footnote{ORCiD IDs: Maximilian Beikirch: 0000-0001-6055-4089, Simon Cramer: 0000-0002-6342-8157, Philipp Otte: 0000-0002-1586-2274, Emma Pabich: 0000-0002-0514-7402, Torsten Trimborn: 0000-0001-5134-7643}, Simon Cramer\footnotemark[1] \footnotemark[2] , Martin Frank\footnote{Karlsruhe Institute of Technology, Steinbuch Center for Computing, Hermann-von-Helmholtz-Platz 1, 76344 Eggenstein-Leopoldshafen, Germany},\\  Philipp Otte\footnote{Forschungszentrum J\"ulich GmbH, Institute for Advanced Simulation, J\"ulich Supercomputing Centre, 52425 J\"ulich, Germany} \footnote{MathCCES, RWTH Aachen University, Schinkelstra\ss e 2,\ 52056 Aachen, Germany} \footnotemark[2], Emma Pabich\footnotemark[1], Torsten Trimborn\footnote{IGPM, RWTH Aachen University, Templergraben 55, 52056 Aachen, Germany} \footnote{Corresponding author: trimborn@igpm.rwth-aachen.de}}
	
\maketitle




\begin{abstract}
We study the qualitative and quantitative appearance of stylized facts in several agent-based computational economic market (ABCEM) models. We perform our simulations with the 
SABCEMM (Simulator for Agent-Based Computational Economic Market Models) tool recently introduced by the authors \citep{SABCEMM}.
\changes{Furthermore, we present novel ABCEM models created by recombining existing models and study them with respect to stylized facts as well. 
This can be efficiently performed by the SABCEMM tool thanks to its object-oriented software design.} The code is available on GitHub
\citep{SABCEMMgithub}, such  that all results can be reproduced by the reader.
\end{abstract}

\include{introduction}

\include{method}

\include{examples}

\include{conclusion}
\section*{Acknowledgement}
Torsten Trimborn gratefully acknowledges support by the Hans-Böckler-Stiftung and the RWTH Aachen University Start-Up grant. 
The work was partially funded by the Excellence Initiative of the German federal and state governments.\\

\appendix

\include{appendix}


	\bibliography{Quellen/SABCEMM.bib}

\end{document}

%% file: introduction.tex
\section{Introduction}
\changes{
Stylized facts are commonly accepted as persistent empirical pattern in financial data.
Probably, the first documented stylized fact is the inequality of income discovered by Pareto in 1897 \cite{pareto1897cours}.
Subsequently, there have been made more statistical observations by Fama and Mandelbrot in the 1960s \citep{mandelbrot1997variation, brada1966letter, FamaPhd}, namely, the non-Gaussian and fat tail behavior of the stock return distribution and volatility clustering.
The non-Gaussian behavior can be conveniently studied with the help of a quantile-quantile plot or the excess kurtosis (see appendix definition \ref{def:exkurt}).
The tail behavior of an empirical distribution is frequently derived by the Hill estimator (see appendix definition \ref{def:hill_estimator}). 
Volatility clustering describes the existence of a positive auto-correlation for squared and absolute stock returns and absence of auto-correlation for raw returns (see appendix definition \ref{def:cor}). 
Nowadays, there have been documented more than 30 stylized facts \citep{chen2012agent, lux2008stochastic}.
For a comprehensive overview of stylized facts, we refer to \citep{cont2001empirical, ehrentreich2007agent, campbell1997econometrics, pagan1996econometrics, lux2008stochastic}. \\[2em]
Standard models in financial literature, such as the capital asset pricing model \citep{sharpe1964capital, lintner1965security}, are build on the efficient market hypothesis by Fama \citep{fama1965behavior}.
Unfortunately, several stylized facts cannot be explained by the efficient market hypothesis and their origin remain unknown \citep{pagan1996econometrics, cowan2002heterogenous, maldarella2012kinetic}.
In the past decades, agent-based computational economic market (ABCEM) models have become a powerful tool in order to generate artificial financial data which feature stylized facts.
The common goal of many ABCEM models is to shed light on the creation of stylized facts.\\[2em]
ABCEM models consider heterogeneous interacting agents which are studied by means of Monte-Carlo simulations.
In contrast to classical financial market models, they share many similarities with interacting particle systems from physics \citep{sornette2014physics, zschischang2001some, lux2008applications}.
Many ABCEM models are able to replicate the most prominent stylized facts of financial markets and are consequently able to provide sufficient conditions for the creation of those.
Examples indicate that behavioral aspects of financial investors play an important role in the creation of stylized facts  \citep{cross2005threshold, lux2008stochastic, chen2012agent}.
Further prominent agent-based models are \citep{kirman1993ants, cont2000herd, lux1999scaling, brock1997rational}.
The evident drawback of this computational approach is that all results are based on numerical experiments only.
Furthermore, this ansatz only constitutes the sufficient conditions for the generation of stylized facts but does not indicate the necessary ones. \\[2em]
Regarding the simulation of agent-based economic market models, there are two major issues present.
First, it has been documented that the appearance of stylized facts are due to finite-size effects as discussed in  \citep{egenter1999finite, zschischang2001some, challet2002, kohl1997influence, hellthaler1996influence}.
In order to exclude such numerical artifacts, it is of major importance to simulate ABCEM models with a large number of agents, which is a computationally expansive undertaking.
Secondly, it is difficult to carry out an objective comparison between different ABCEM models, since the models are implemented in different languages and simulated on different hardware.
Besides, we experienced difficulties while reproducing the results published in literature.
This may well be due to the sensitivity of ABCEM models to their parameters and due to sometimes incomplete information in publications regarding details of the implementation, such as initial values of model quantities. \\[2em]
Therefore, we intend to reproduce established results of multiple models on an identical software architecture to permit an objective comparison.
{In particular, we investigate the Levy-Levy-Solomon (LLS) model \citep{levy1994microscopic}, the Cross model \citep{cross2005threshold} and the Franke-Westerhoff model \citep{franke2012structural} regarding their long time behavior or their finite size effects.}\\
Finally, we create novel ABCEM models and study them with respect to stylized facts.
These new models are built out of well known building blocks, such as a specific market mechanisms or agent designs of other ABCEM models.
This approach allows the study of the interplay of different modeling aspects on the creation of stylized facts. \\[2em]
For our simulations, we utilize the recently introduced open source SABCEMM simulator \citep{SABCEMM} available on GitHub \citep{SABCEMMgithub}. There are two major reasons for that choice: 
First, the SABCEMM simulator is efficient in the sense that it is allows for simulating ABCEM models with up to several million agents on a standard notebook.
Secondly, the simulator is build on the idea of  building blocks such as market mechanisms or agent designs and thus supports the easy recombination of building blocks of different models as easily as plugging together pieces of a puzzle.}\\[2em]
The outline of the paper is as follows: In section \ref{sabcemm} we give a short introduction to the SABCEMM simulator. 
Then we present the simulation results of each model with respect to the reproducibility of the most prominent stylized facts, namely fat-tails, absence of auto-correlation and volatility clustering.
Furthermore, we create new models out of existing building blocks of known ABCEM models. Finally, we test these new models with respect to stylized facts. We finish this paper with  a short conclusions of this work.

%% file: method.tex
\section{The SABCEMM Simulator}\label{sabcemm}

\begin{figure}[h!]
\begin{center}
\includegraphics[width=0.45\textwidth]{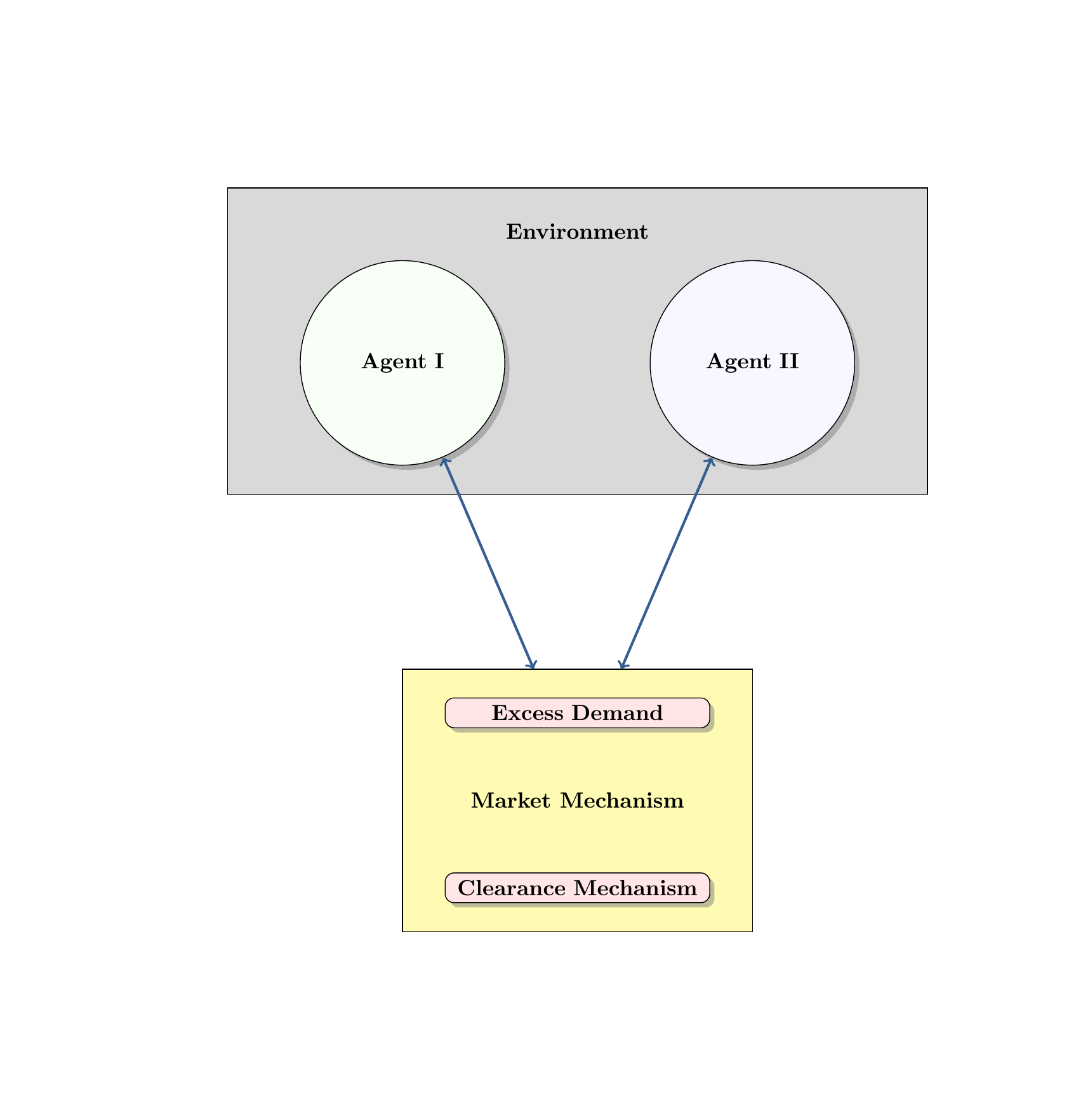}
\caption{Schematic picture of the abstract ABCEMM model \citep{SABCEMM}. }\label{OurModel}
\end{center}
\end{figure}

The recently introduced open source simulator SABCEMM \citep{SABCEMM} is designed especially for large-scale simulations of ABCEM models.
\changes{This simulator implements an object oriented design leveraging a generalized structure of ABCEM models as defined in \citep{SABCEMM}.
The implementations of the individual ABCEM building blocks are well-separated and the ABCEM model is assembled from the building blocks via an XML-based configuration file.
Hence, the evaluation of an ABCEM model using a different building block, such as the market mechanism, requires only a change of the configuration file.
If the changed building block does not already exist only this single block has to be implemented.
In the following, we present the main conceptual ideas behind the simulator.
\\ \\
SABCEMM is well suited for any economic market model which consists of at least one \textit{agent} and one \textit{market mechanism}.}
An agent is an investor who has a supply of or demand for a certain good or asset, which is traded at the market. 
The market mechanism determines the price from the demand and supply of all market participants.
More precisely, we differentiate between the so called \textit{price adjustment process} and the \textit{excess demand calculator}.
The latter one aggregates the supply and demand of all market participants (agents) to a single quantity, the excess demand.
The former one represents the method of how the market price is fixed based on this excess demand.
A schematic picture, which illustrates the presented ideas is shown in Figure \ref{OurModel}.  
The concept of an \textit{environment} has been first introduced by the authors in \citep{SABCEMM}.
An environment represents possible additional coupling between the agents.
Probably, the most famous example for an environment is herding, which is frequently used in ABCEM models.
We emphasize that such an environment is not mandatory.
For a rigorous mathematical definition of the meta-model which is the foundation of the SABCEMM simulator and a detailed discussion of technical details and computational aspects of SABCEMM, we refer to \citep{SABCEMM}.

%% file: examples.tex
\section{Validation and Novel Tests for Known ABCEM Models}

\label{examples}
\changes{In this section we present simulations of the Cross, LLS and Franke-Westerhoff models.}
We discuss each model separately and demonstrate the advantages of our simulation framework. 
We study the behavior of the LLS and Cross models for large numbers of agents to investigate if the models exhibit finite-size effects.
Furthermore, we study several model variants of the Franke-Westerhoff model with respect to common stylized facts. 
We emphasize that we provide all necessary information to reproduce the results.
To allow for reproducibility, we define the implemented models in detail and present all parameter values and initial states for each model in the appendix \ref{appendixModel}.
We ran our simulations on an Intel Xeon 64 bit architecture.
The input files for all simulations can be found on the GitHub repository of the SABCEMM source code \citep{SABCEMMgithub}.
Additionally, our simulation data is available as a data publication \citep{DataS, DataSet2}.\\ \\
We provide a short introduction to each model and discuss the appearance of stylized facts for each model separately.
For a detailed definition of the implemented models and parameter settings we refer to appendix \ref{appendixModel}.

\subsection{Cross Model}
\label{OM-Cross}
\begin{figure}[h!]
	\begin{subfigure}{\textwidth}
	    \begin{center}
	        \includegraphics[width=0.7\textwidth]{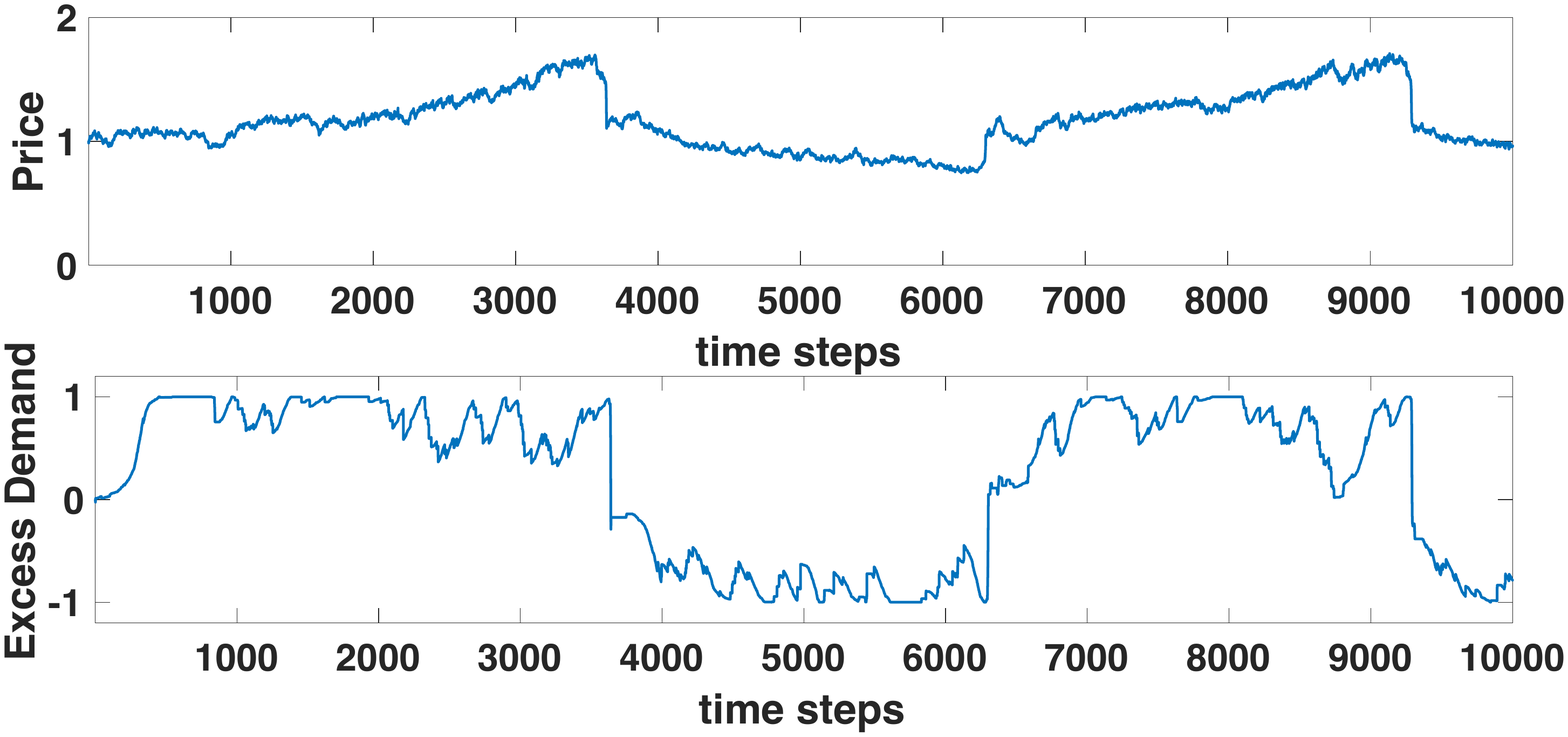}
	        \caption{Parameters as in \cref{cross-basic-parameter}.}
	        \label{OM-Cross-Base}
	    \end{center}
	\end{subfigure}
	\begin{subfigure}{\textwidth}
	    \begin{center}
	        \includegraphics[width=0.7\textwidth]{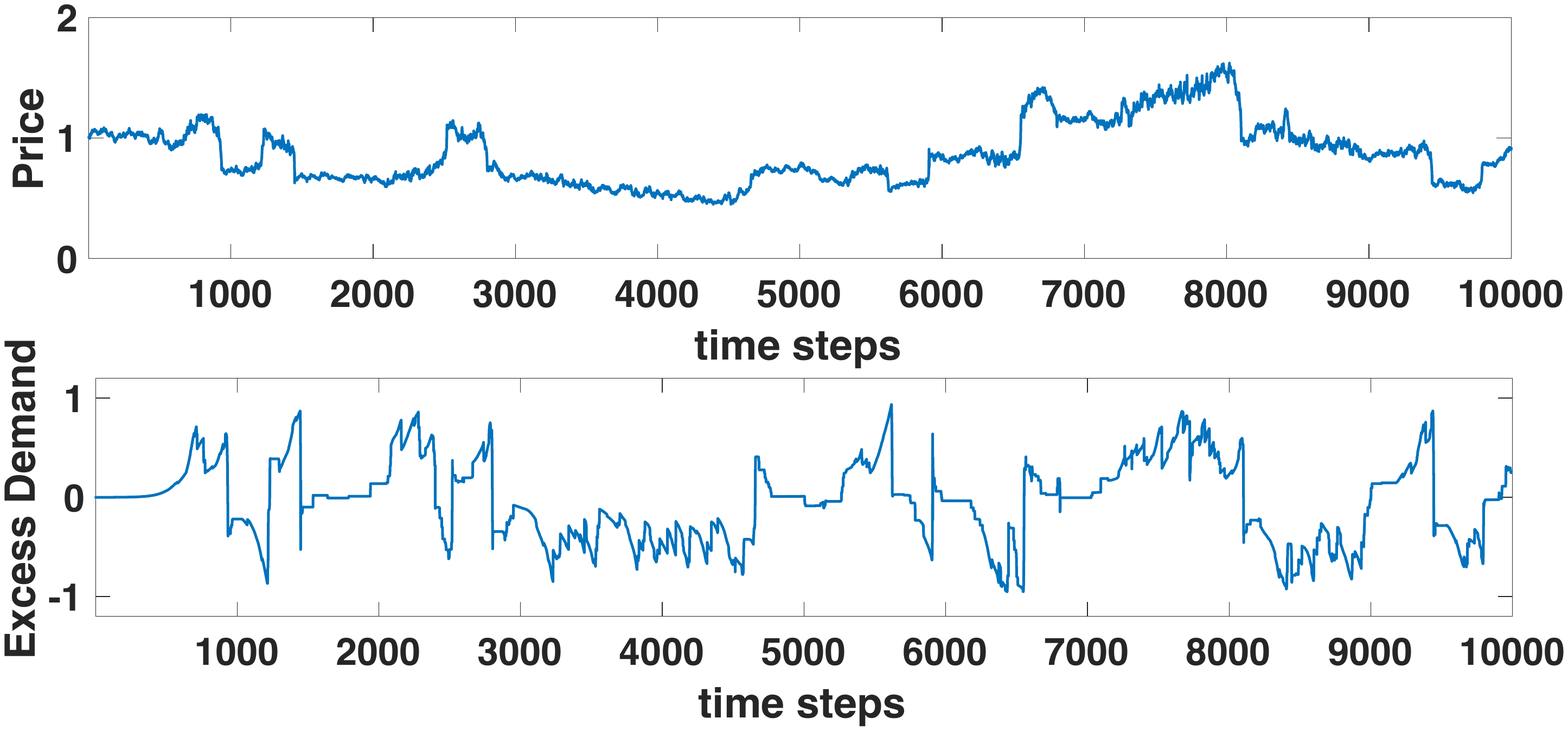}
	        \caption{Parameters as in \cref{cross-basic-parameter}, except $N=5,000,000$ and $\theta = 2$.}
	        \label{MA-Cross-Hetero}
	    \end{center}
	\end{subfigure}
	\caption{Development of price and excess demand for the Cross base model.}
\end{figure}

\begin{figure}[h!]
	\begin{subfigure}{\textwidth}
	    \begin{center}
	        \includegraphics[width=0.7\textwidth]{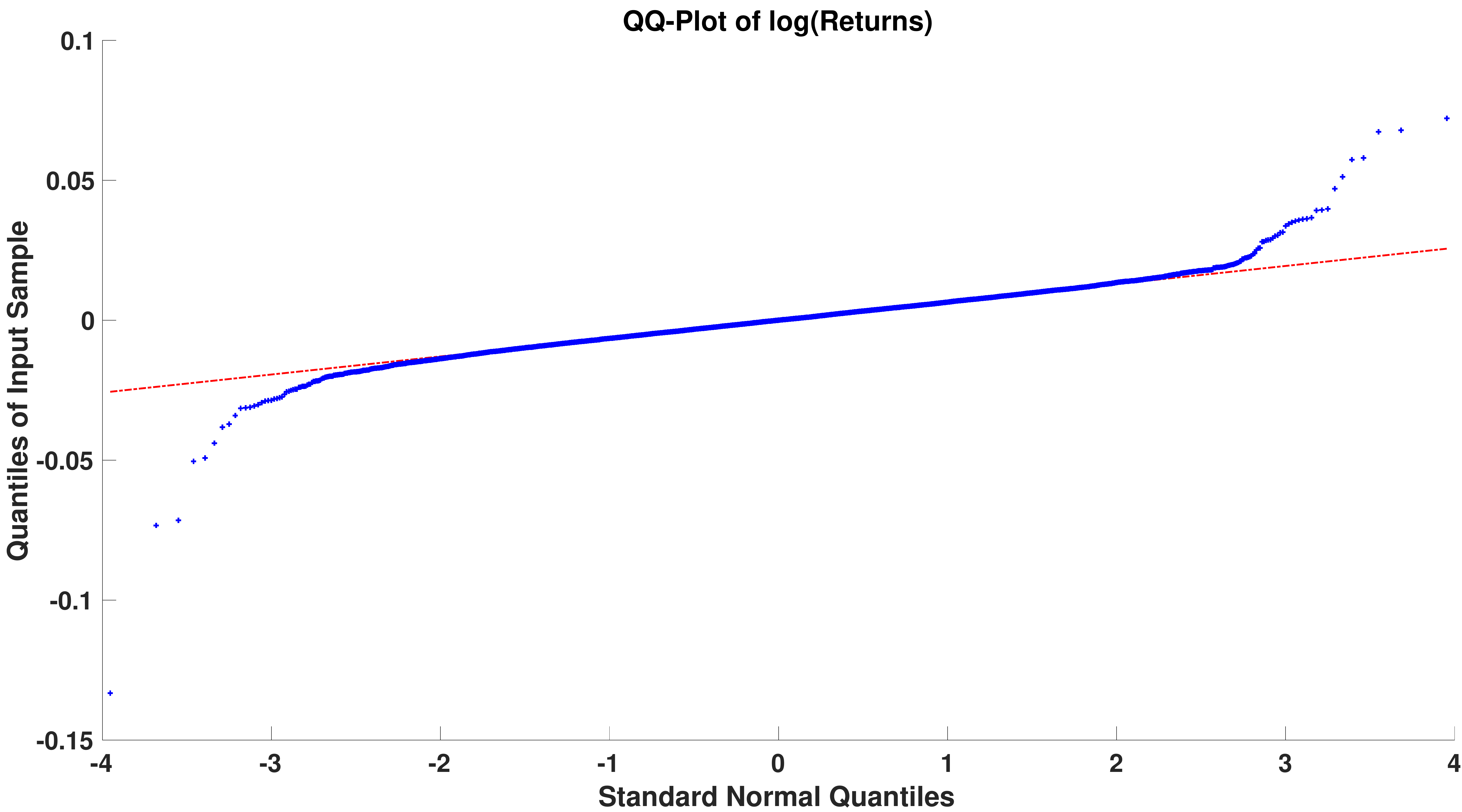}
	        \caption{Parameters as in \cref{cross-basic-parameter}.}
	        \label{OM-Cross-Base-qq}
	    \end{center}
	\end{subfigure}
	\begin{subfigure}{\textwidth}
		\begin{center}
	        \includegraphics[width=0.7\textwidth]{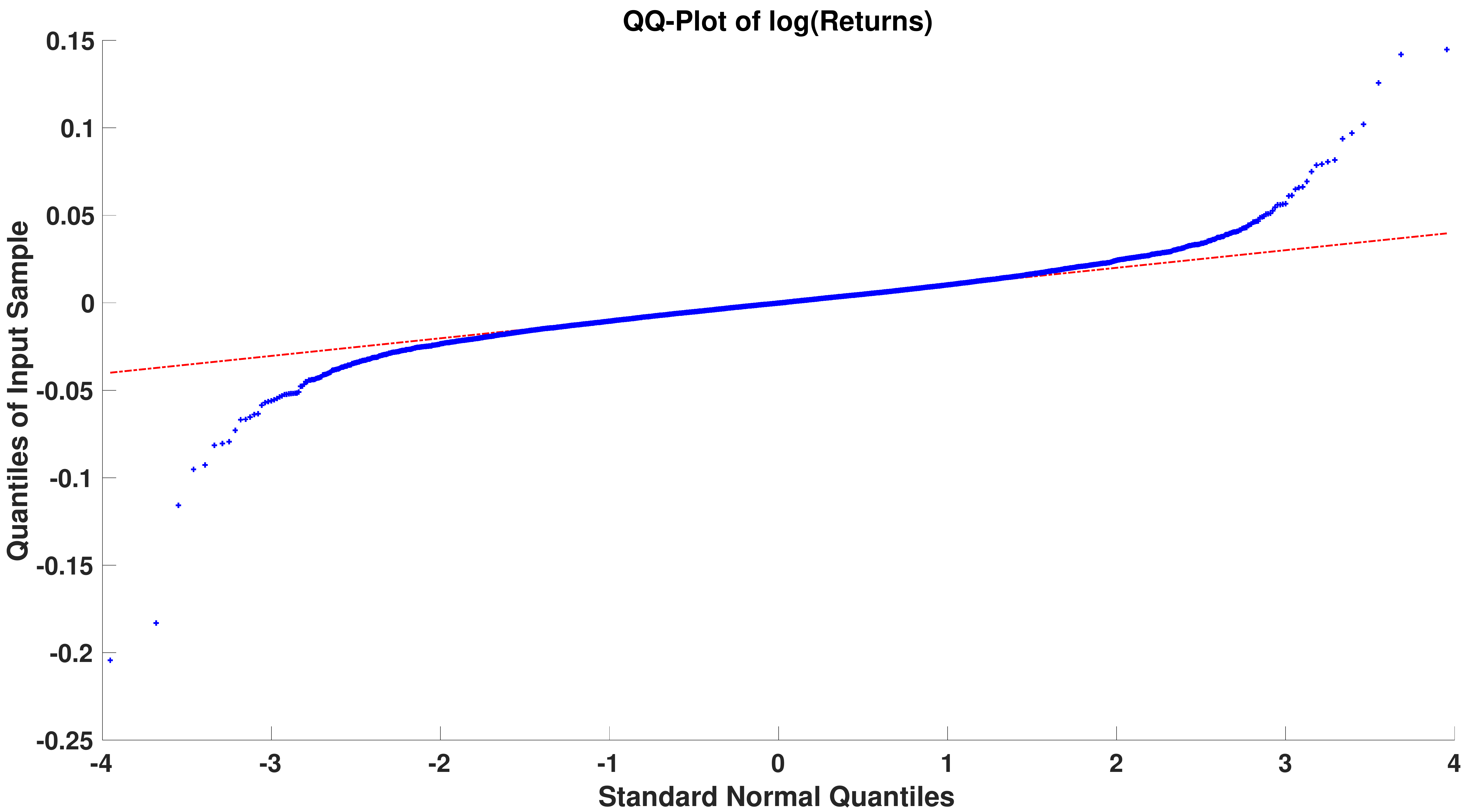}
	        \caption{Parameters as in \cref{cross-basic-parameter}, except $N=5,000,000$ and $\theta = 2$.}
	        \label{MA-Cross-Hetero-qq}
	    \end{center}
	\end{subfigure}
	\caption{Fat-tails observed in QQ-plots of the logarithmic returns for the Cross base model.}
\end{figure}

\begin{figure}[h!]
	\begin{subfigure}{\textwidth}
	    \begin{center}
	        \includegraphics[width=0.7\textwidth]{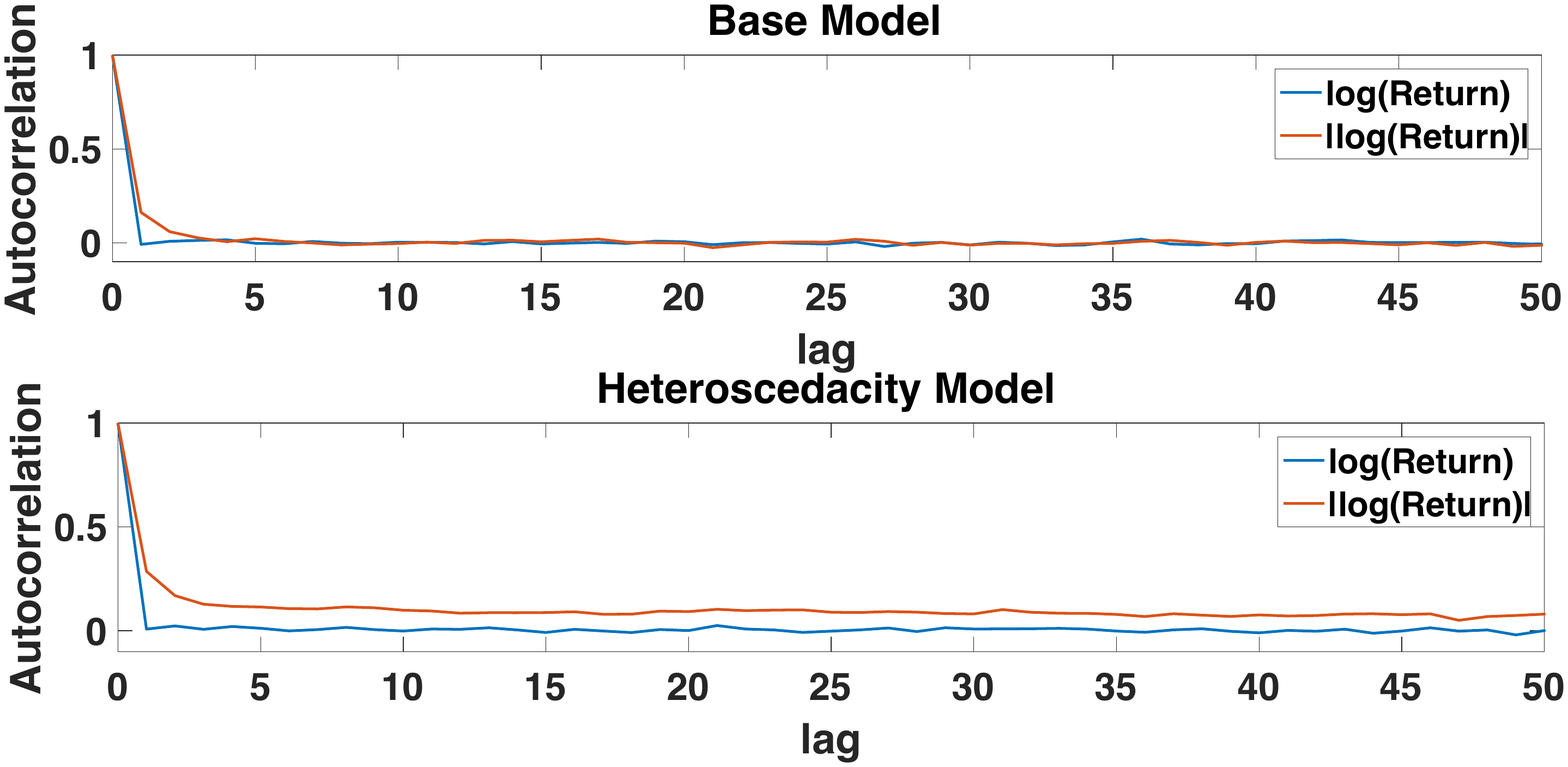}
	        \caption{Upper graph: Cross base model with parameters see \cref{cross-basic-parameter}; lower graph: Cross heteroscedacity model with parameters see \cref{cross-basic-parameter} except $\theta =2$.}
	        \label{OM-Cross-AutoCorr}
	    \end{center}
    \end{subfigure}
    \begin{subfigure}{\textwidth}
	    \begin{center}
	        \includegraphics[width=0.7\textwidth]{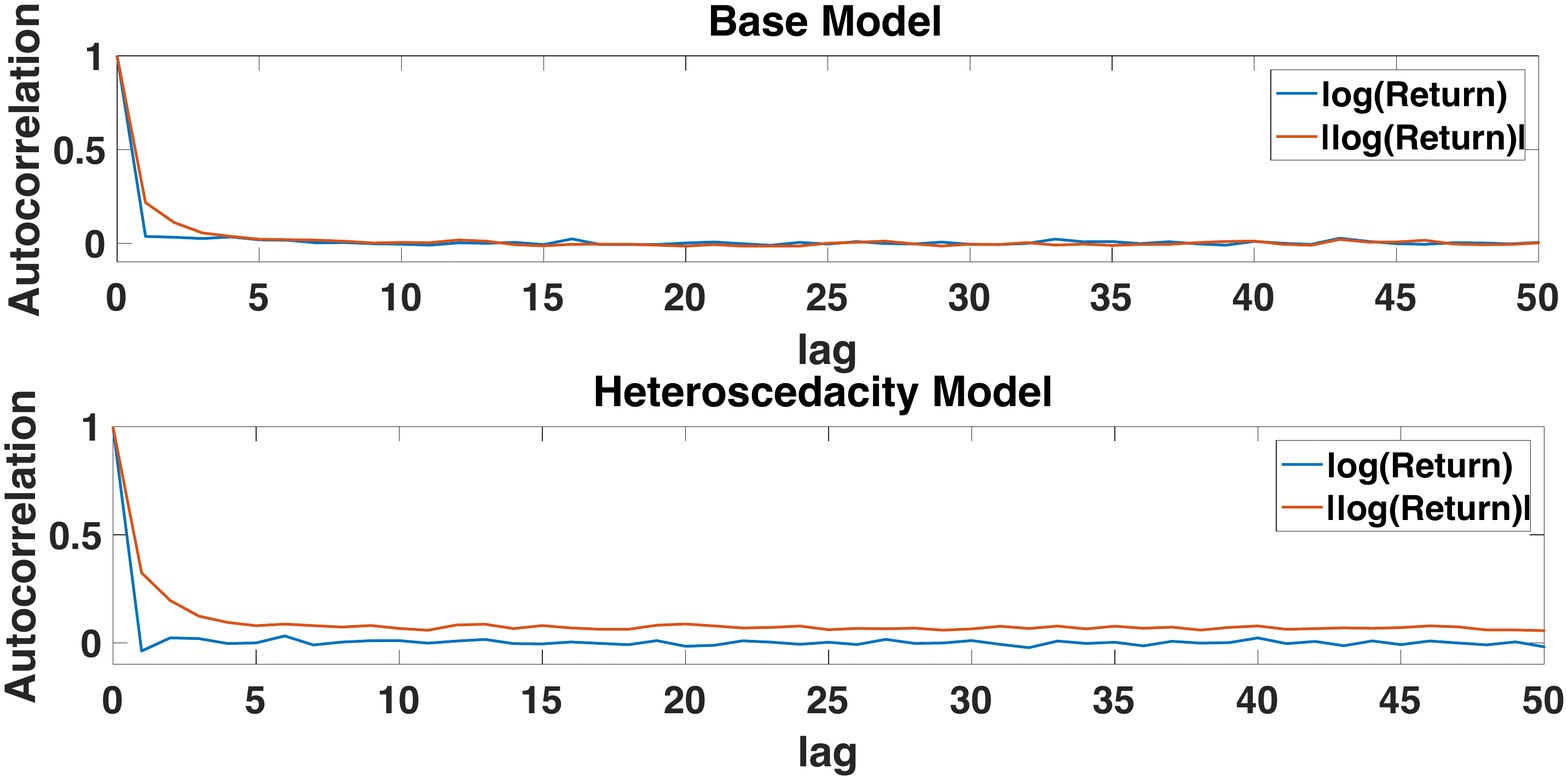}
	        \caption{Upper graph: Cross base model with parameters see \cref{cross-basic-parameter} and $N=5,000,000$; lower graph: Cross heteroscedacity model with parameters see \cref{cross-basic-parameter} except $\theta =2$ and $N = 5,000,000$.}
	        \label{MA-Cross-AutoCorr}
	    \end{center}
    \end{subfigure}
    \caption{Auto-correlation of log-returns and absolute log-returns in the Cross base model and the heteroscedacity models.}
\end{figure}

This section presents results for the Cross model which is inspired by the Ising model \citep{ising1925beitrag} from physics.
In the Cross model, each agent is characterized by his position, long $\sigma_i=+1$ or short $\sigma_i=-1$ in the market, respectively.
Their investment propensity is determined by two tensions: one related to rational agent behavior and the other to irrational agent behavior. 
They both mimic the role of temperature in the Ising model.
The irrational agent behavior takes into account the herding propensity of financial investors. 
The price process is driven through the change of the excess demand and is additionally perturbed by white noise. 
The authors Cross et al. show that their model can replicate the most prominent stylized facts of financial markets, namely fat-tails, uncorrelated price returns and volatility clustering.
For further modeling details, we refer to \citep{cross2005threshold, cross2007stylized}.\\ \\
In our simulations, we obtain the same qualitative results as presented in \citep{cross2005threshold, cross2007stylized}.
As Figure \ref{OM-Cross-Base} reveals, the price dynamic is influenced heavily by the evolution of the excess demand over time.
Furthermore, the absence of auto-correlation in  raw price returns can be verified by Figure \ref{OM-Cross-AutoCorr}.
In addition, Figure \ref{OM-Cross-Base-qq} reveals the fat-tail in asset returns.
By adding the heteroskedasticity parameter $\theta$, one couples the noise with the excess demand.
This leads to volatility clustering as we can see in Figure \ref{OM-Cross-AutoCorr}. 
A detailed introduction of the model can be found in appendix \ref{appendixModel}.

\paragraph{Finite Size Effects}

In \citep{cross2007stylized} the authors claim that their model has no finite size effects.
Despite of their claim, all their simulations are performed with 100 agents only.
In order to verify their statement, we analyze the model with different numbers of up to five million agents. We ran our simulations with $10,000$ time steps in order to have a sufficiently large sample size. Our simulations support the findings of Cross et al. (see \cref{MA-Cross-Hetero,MA-Cross-Hetero-qq,MA-Cross-AutoCorr}).
Hence, the qualitative behaviour of the Cross model is insensitive to the number of agents, be it $100$ or five million agents.


\subsection{LLS Model}
\label{OM-LLS}
\begin{figure}[h!]
    \begin{center}
        \includegraphics[width=0.7\textwidth]{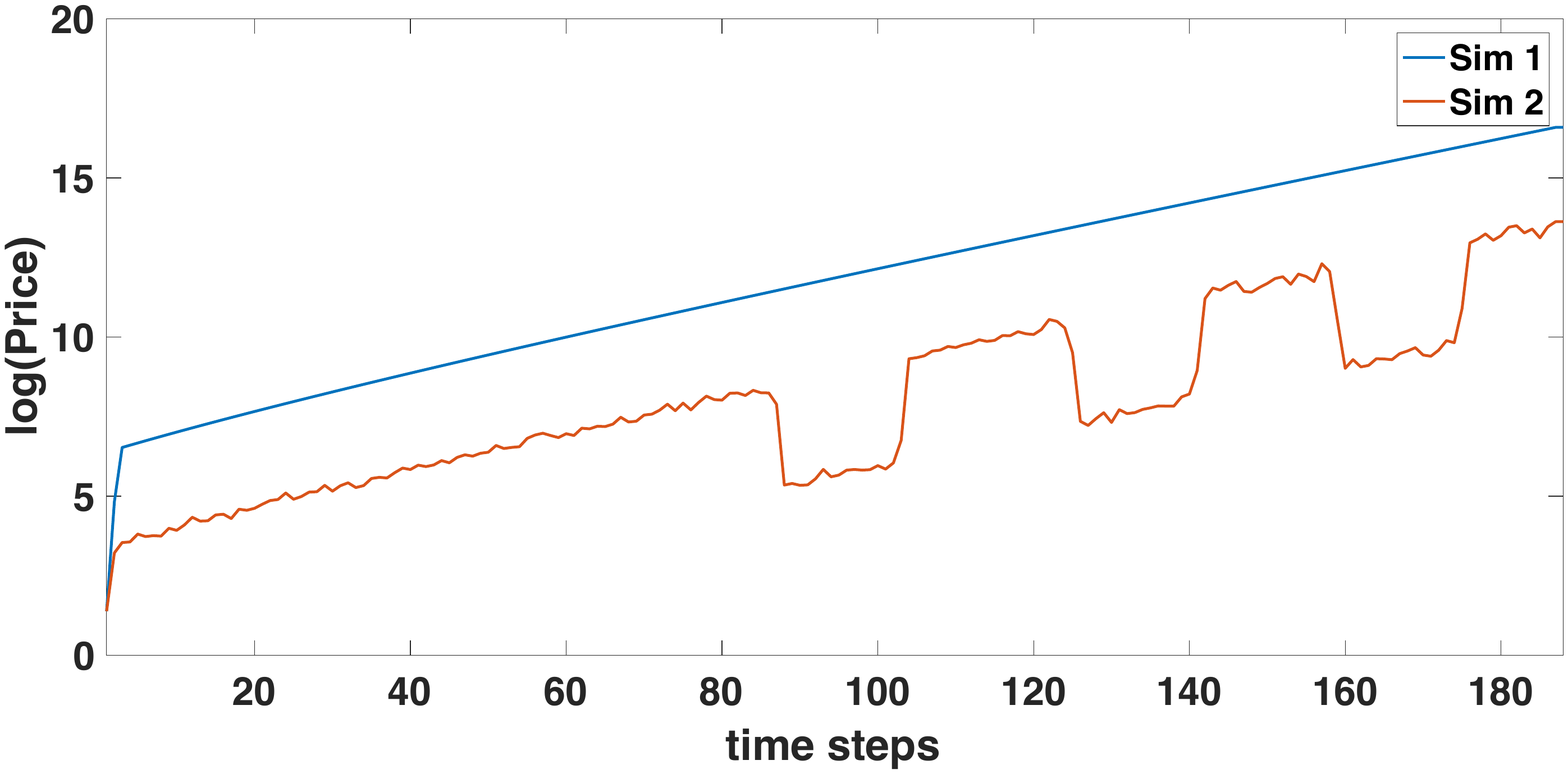}
        \caption{Price evolution of the LLS model with noise $\sigma_{\gamma}=0.2$ (red) and without noise $\sigma_{\gamma}=0$ (blue). Parameters as in \cref{LLS-basic}}
        \label{OM-LLS-1+2}
    \end{center}
\end{figure}

\begin{figure}[h!]
    \begin{center}
        \includegraphics[width=0.7\textwidth]{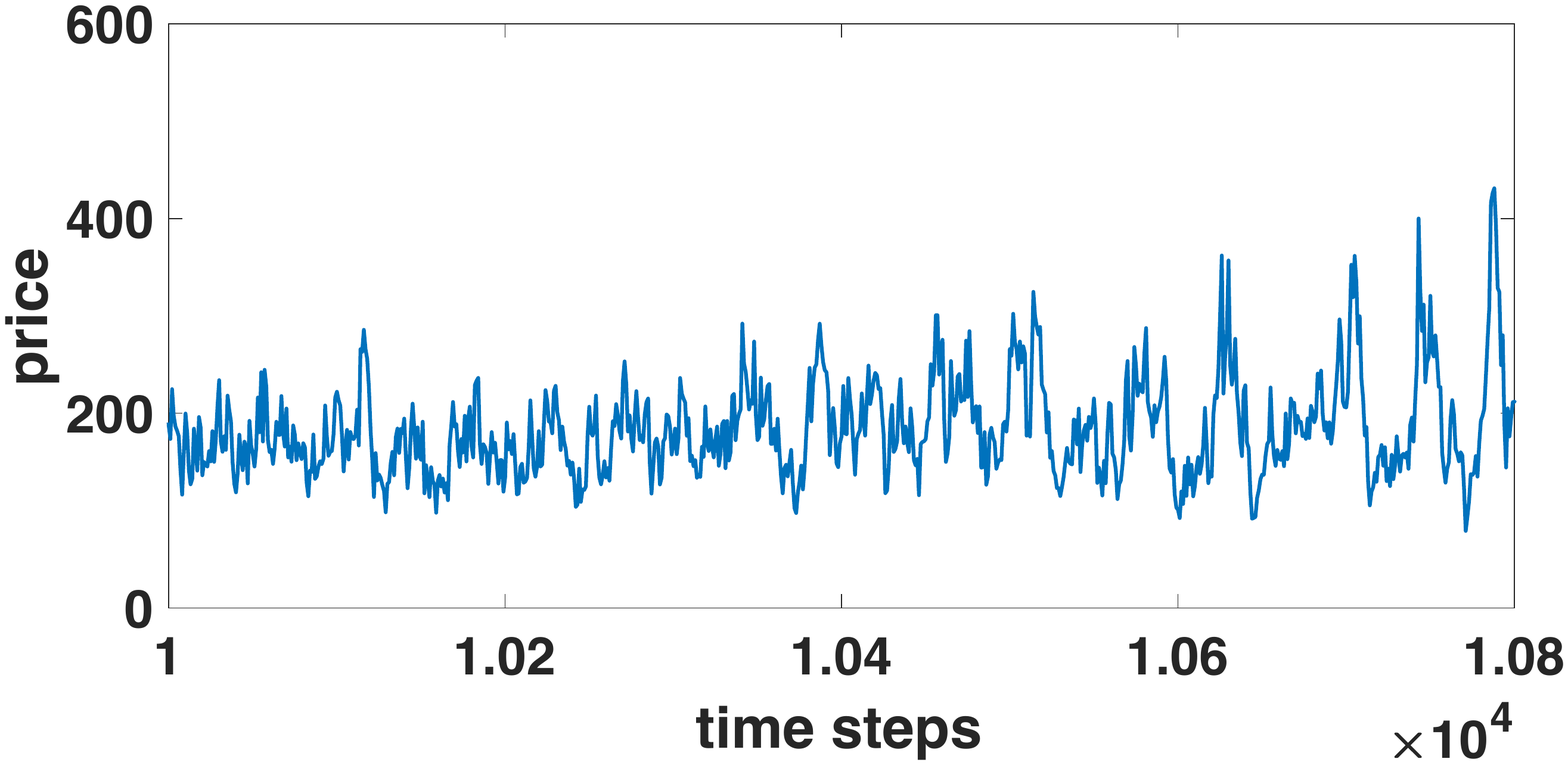}
        \caption{Price evolution of the LLS model with three different investor types. Parameters as in \cref{LLS-3-agents}. }
        \label{OM-LLS-3}
    \end{center}
\end{figure}

In this section, we present results for the LLS model which is one of the earliest and most influential econophysical ABCEM models.
In addition, the LLS model is an example of a model assuming a rational market, i.e. \changes{the price equilibrates at each time step and the excess demand tends to zero} (compare \cref{fixedpointLLS}).
For further details we refer to \citep{SABCEMM}.
Note that the LLS model is subject to critical discussions in literature \citep{zschischang2001some}.
We discuss these crucial findings in detail in this section.\\ \\
The LLS model considers the wealth evolution of the financial agents.
Every agent has to decide in each time step which fraction of wealth he wants to invest in stocks with the remaining wealth being invested in a safe bond.
The investment decision is determined by a utility maximization.
For modeling details we refer to \citep{levy1994microscopic, levy1995microscopic, levy1996complex, levy2000microscopic}.\\ \\
We define the model and parameter sets in detail in appendix \ref{appendixModel}, which is the identical choice as in the earlier studies \citep{ levy1995microscopic, levy1996complex}.
We consider only one type of financial agent. 
Figure \ref{OM-LLS-1+2} shows the simulation in the case of noise and no noise added to the investment decision for an agent with fixed memory span of $m=15$.
We observe that the noise leads to oscillatory behavior, which coincides with earlier findings in \citep{levy1995microscopic}.
Figure \ref{OM-LLS-3} shows results for three types of agents with different memory spans $m_1=10,\ m_2=141,\ m_3=256$.
The results are qualitatively identical to the results in \citep{levy1996complex}.  

\begin{figure}[h!]
    \begin{center}
        \includegraphics[width=0.49\textwidth]{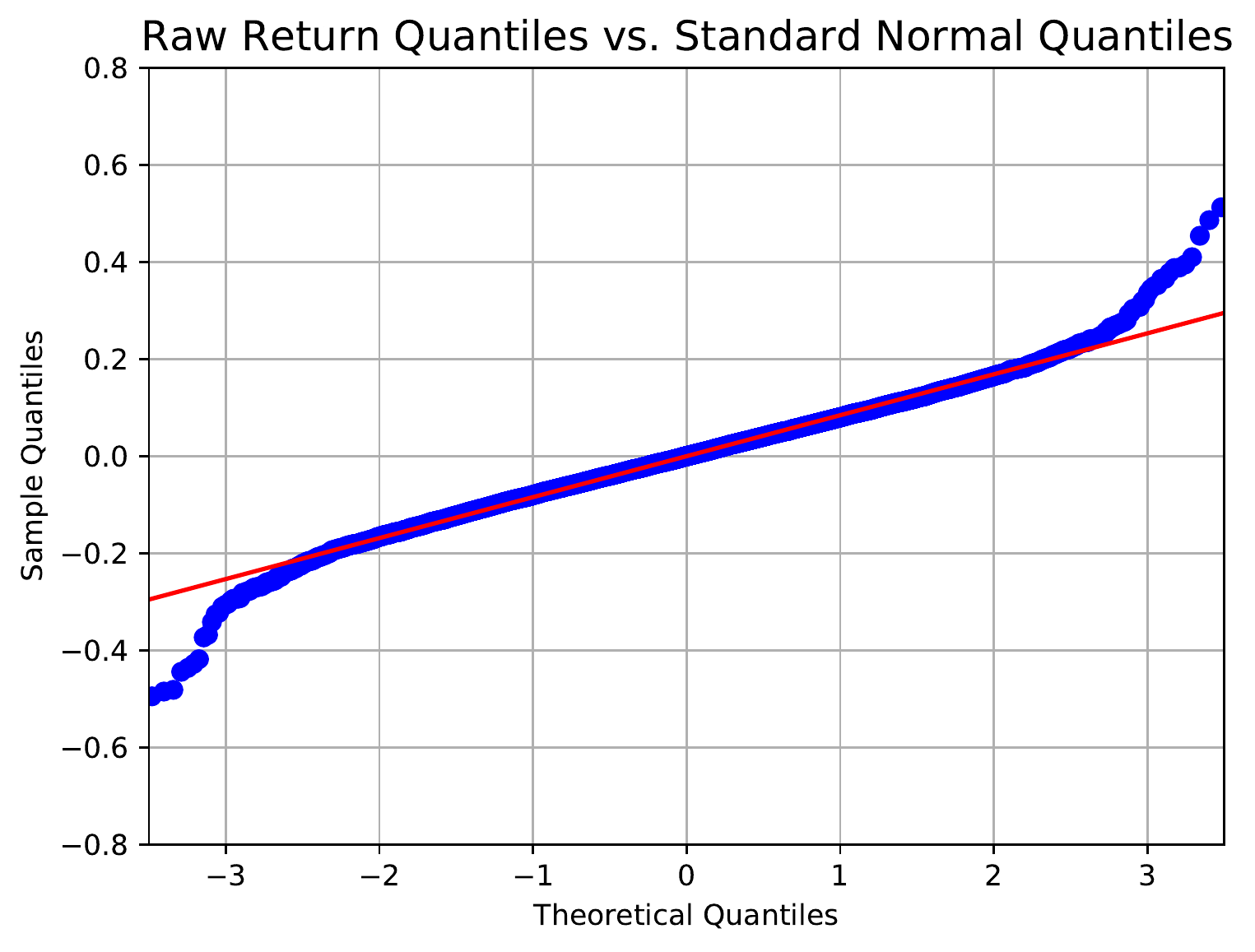}
        \includegraphics[width=0.49\textwidth]{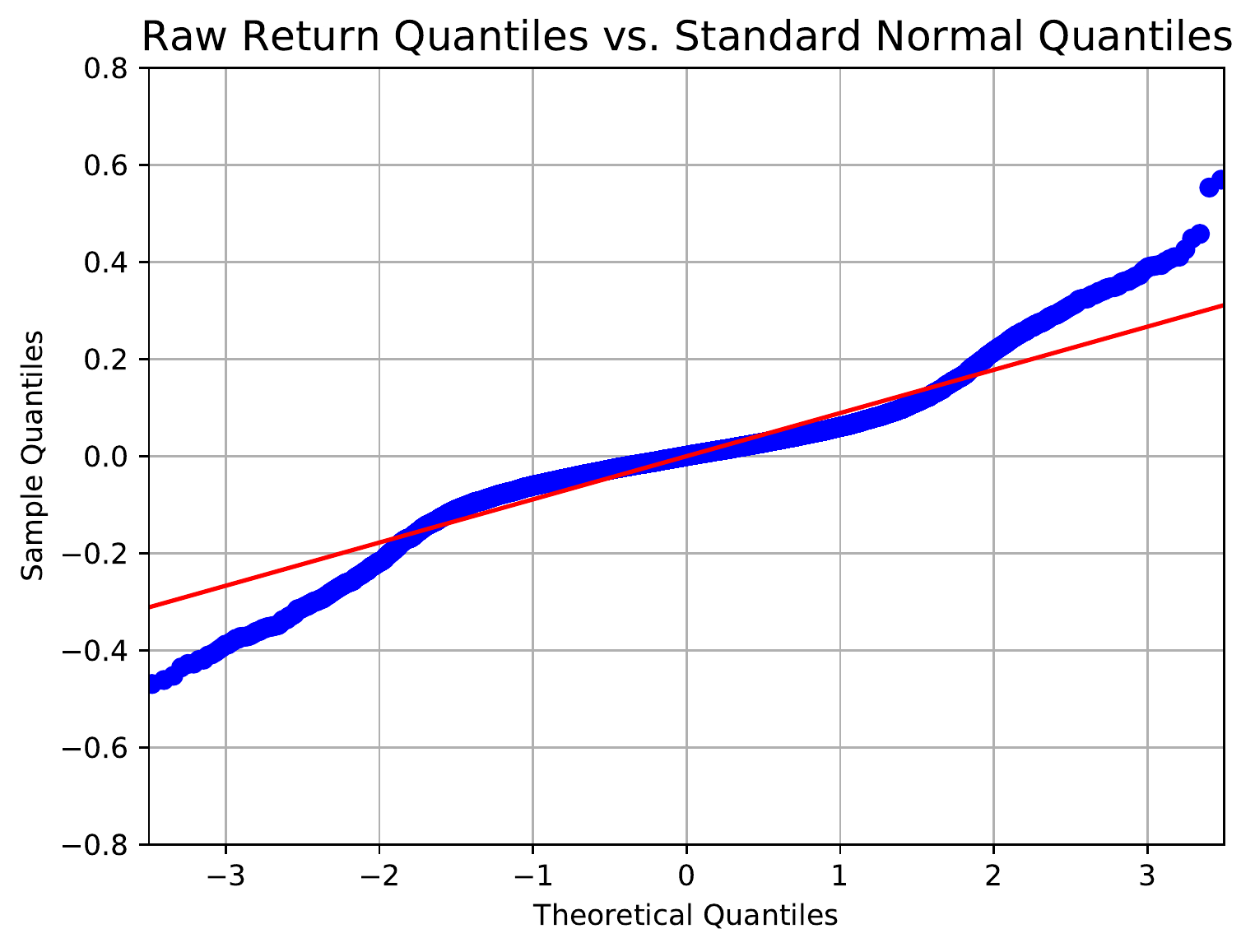}
        \caption{QQ-Plot of log-returns. Simulations conducted with $99$ agents (left) and $999$ agents (right), respectively. The parameters are set as in table \ref{LLS-3-agents}.}
        \label{OM-LLS-3-4-qq}
    \end{center}
\end{figure}

\begin{figure}[h!] 
	\begin{subfigure}{\textwidth}
	    \begin{center} 
	        \includegraphics[width=0.7\textwidth]{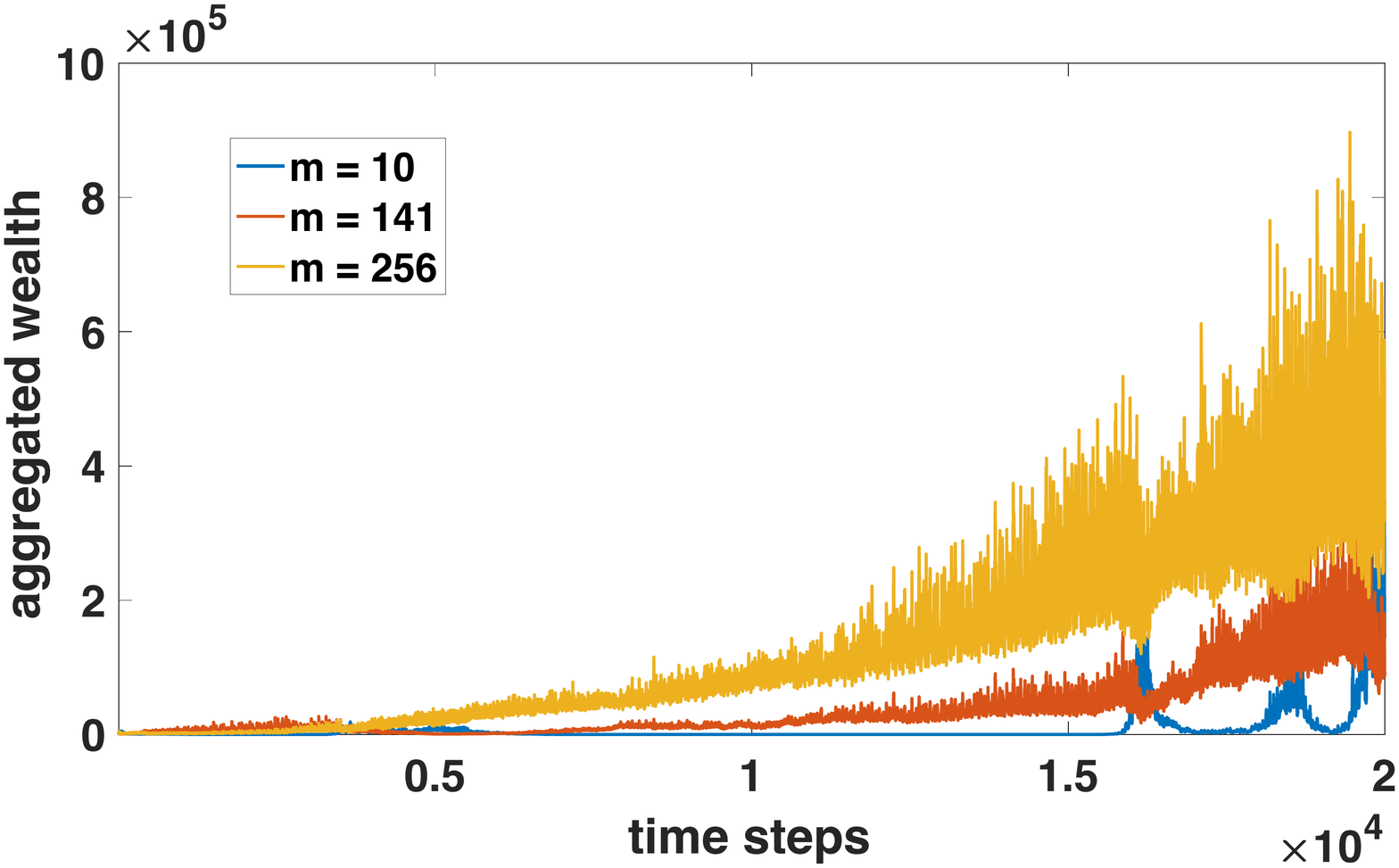} 
	        \caption{Results for $N=99$ agents.} 
	        \label{MA-LLS-cash1} 
	    \end{center} 
	\end{subfigure}
	\begin{subfigure}{\textwidth}
	    \begin{center} 
	        \includegraphics[width=0.7\textwidth]{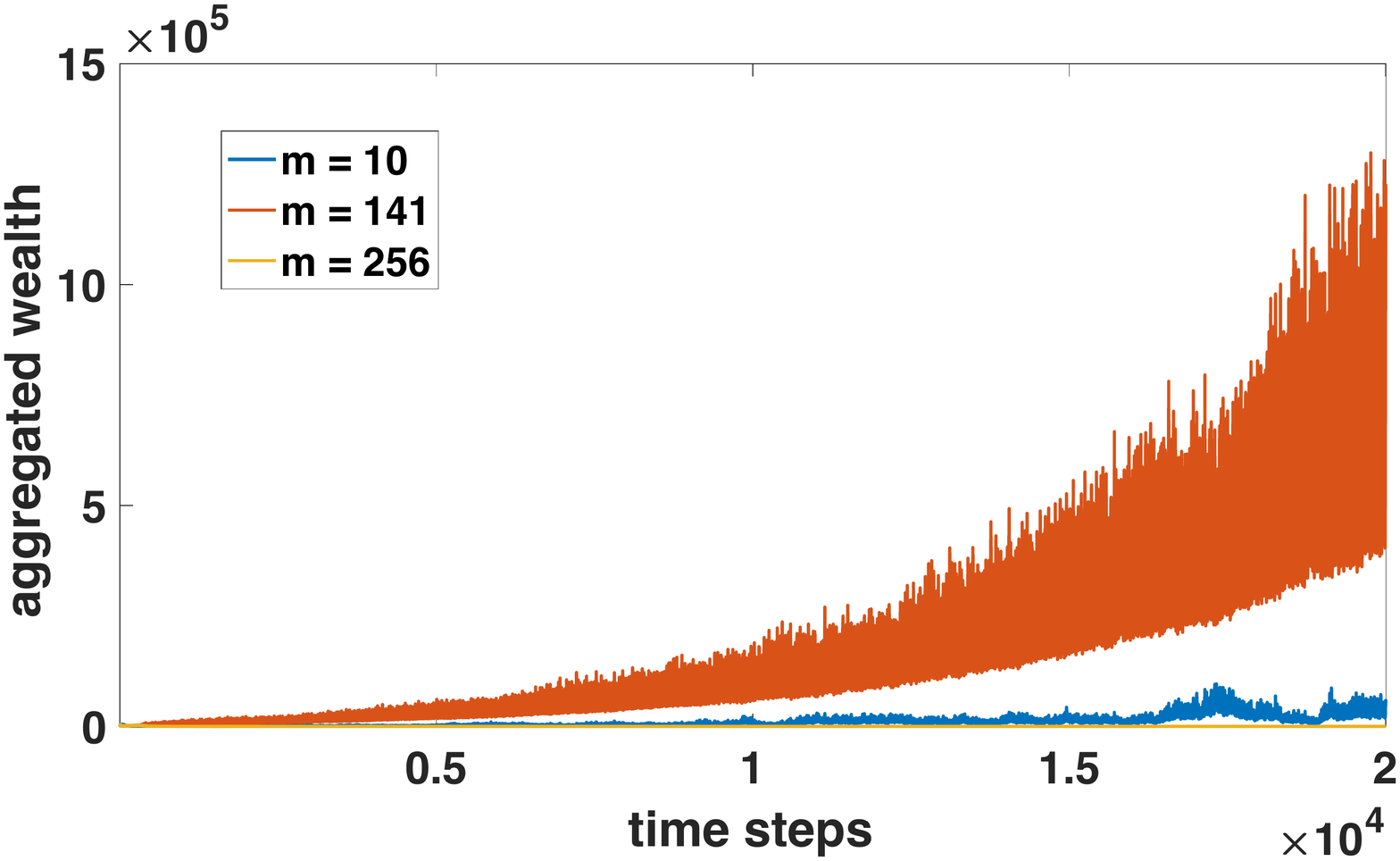} 
	        \caption{Results for $N=999$ agents.} 
	        \label{MA-LLS-cash3} 
	    \end{center} 
	\end{subfigure}
	\caption{Aggregate wealth of three equally sized agent groups for different total numbers of agents and parameters as found in table \ref{LLS-3-agents}.}
\end{figure}

\paragraph{Finite Size Effects}
\label{MA-LLS}
It was discovered earlier that the LLS model exhibits finite size effects \citep{zschischang2001some}.
In our simulations, we identify two different kinds of effects caused by a different number of agents.
Our simulations are conducted first with 99 agents and then with 999 agents. 
First, the qq-plot of the logarithmic stock return of both simulations depicted in Figure \ref{OM-LLS-3-4-qq} clearly shows that the number of agents has a tremendous effect on the tail behavior.
Secondly, Levy at al. \citep{levy1996complex} claimed that the investor group with the maximum memory becomes the dominant one, meaning they own the maximum amount of wealth. 
In our simulations, the wealth evolution of different agent groups changes with varying number of agent, as the Figures \ref{MA-LLS-cash1} and \ref{MA-LLS-cash3} reveal.  
Thus, we can conclude that the qualitative output of the model changes with respect to the number of agent, which is an undesirable model characteristic.  

\begin{figure}[h!]
    \begin{center}
        \includegraphics[width=\textwidth]{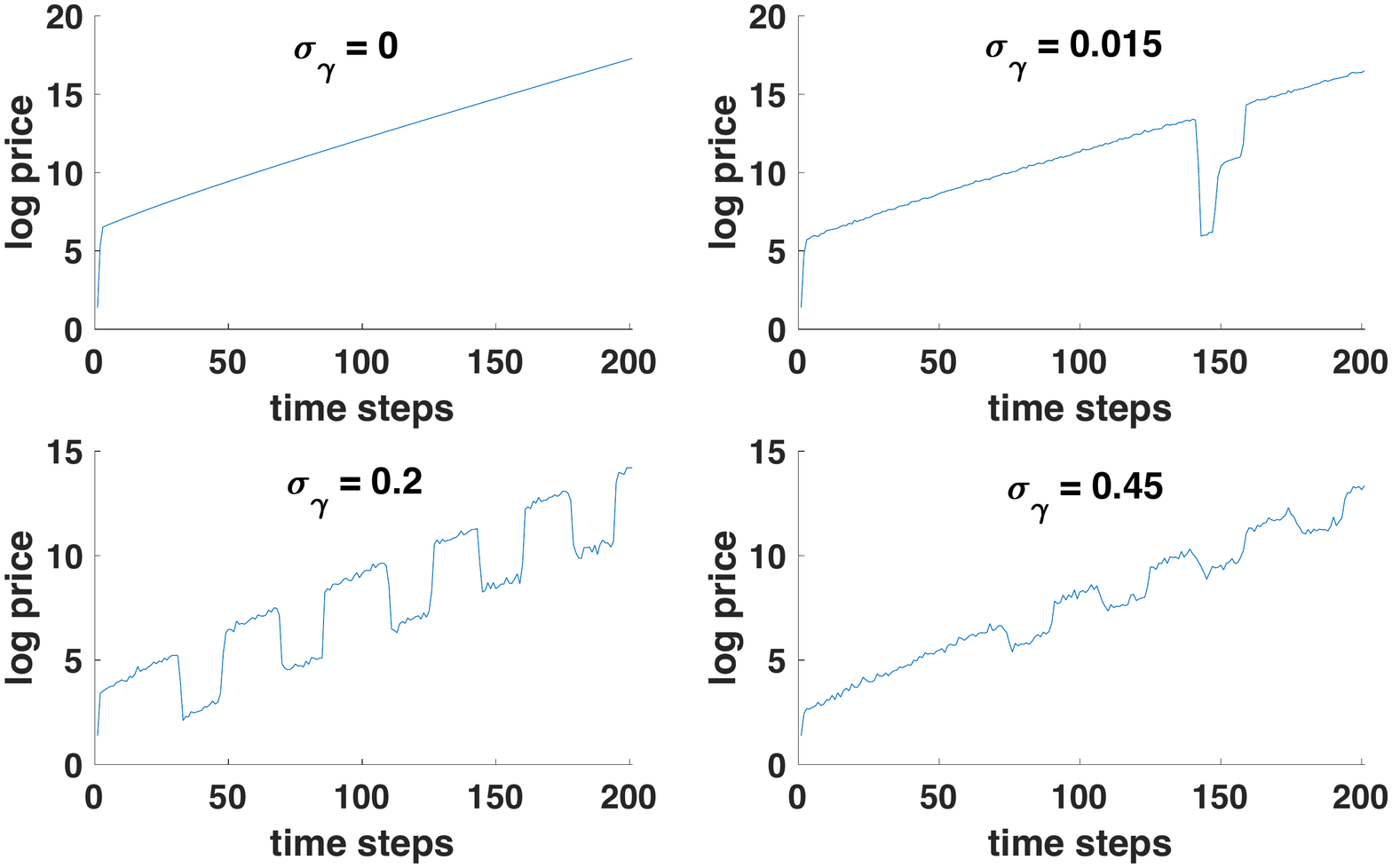}
        \caption{Simulations of the basic LLS model with varying noise levels. Parameters as set in table \ref{LLS-basic}.}
        \label{LLS-NoiseLevel}
    \end{center}
\end{figure}

\paragraph{Discussion of Model Behavior}
The simulation in Figure \ref{OM-LLS-1+2} reveals that the deterministic model is characterized by a constant investment proportion. 
The optimal investment proportion is always located at the boundaries $\gamma\in \{0.01, 0.99\}$, determined through the initialization of the return history.
This is an absolutely reasonable result, thus the wealth evolution in the original LLS model \citep{levy1994microscopic} is linear
\begin{align}
w(t_{k+1})=w(t_k) + (1-\gamma(t_k))\ r + \gamma(t_k)\ w(t_k) \ \frac{S(t_{k+1})-S(t_k) +D(t_k)}{S(t_{k})}, \label{LLSwealth}
\end{align}
and the chosen logarithmic utility function is monotonically increasing.
In fact, additive noise on the optimal solutions leads to oscillatory behavior.
Thus, the investors change between the two possible extreme investments of being fully invested in stocks or bonds.
We point out that the noise level is crucial in order to obtain this oscillatory behavior.
Figure \ref{LLS-NoiseLevel} illustrates the model output for different noise levels. 

Nevertheless Figure \ref{OM-LLS-3} seems to indicate chaotic price behavior.
The previous simulations (see \cref{OM-LLS-3-4-qq,MA-LLS-cash1,MA-LLS-cash3}) clearly reveal  finite size effects.
In our simulations we obtain that also in the noisy case approximately $90\%$ of the investment decisions (pre-noise) are located at the boundaries.
Mathematically this is an unsatisfying result since in the expansive optimization process in useless in $90\%$ of the cases.
Thus, we may conclude that the LLS model exhibits several undesirable model characteristics.

\subsection{Franke-Westerhoff Model}

\begin{figure}[ht]
\centering
\begin{subfigure}{0.49\textwidth}
  \centering
  \includegraphics[width=\linewidth]{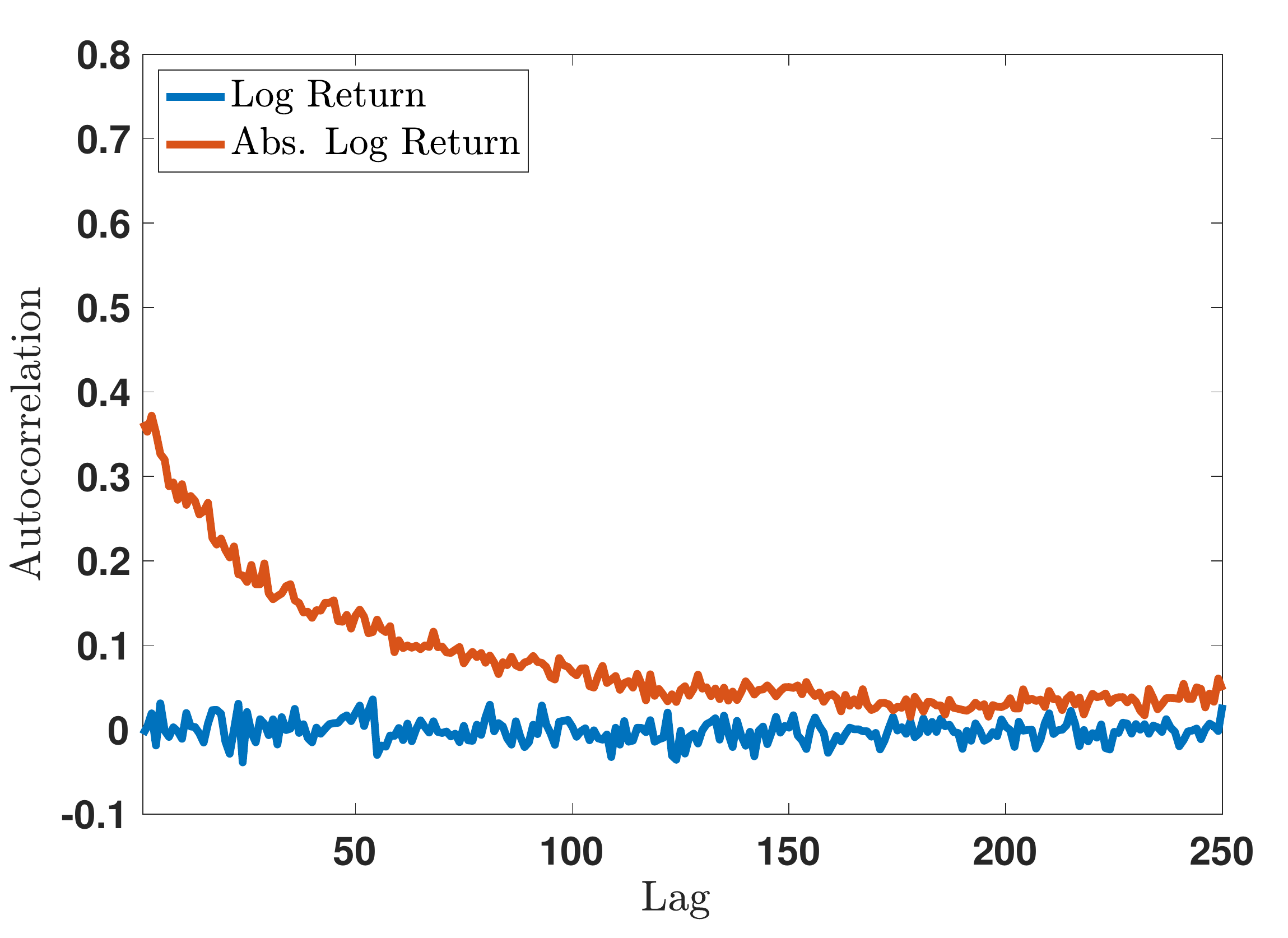}
  \caption{Auto-correlation of logarithmic returns and absolute logarithmic returns.}
  \label{fig:fw_dcahpm_auto}
\end{subfigure}
\begin{subfigure}{0.49\textwidth}
  \centering
  \includegraphics[width=\linewidth]{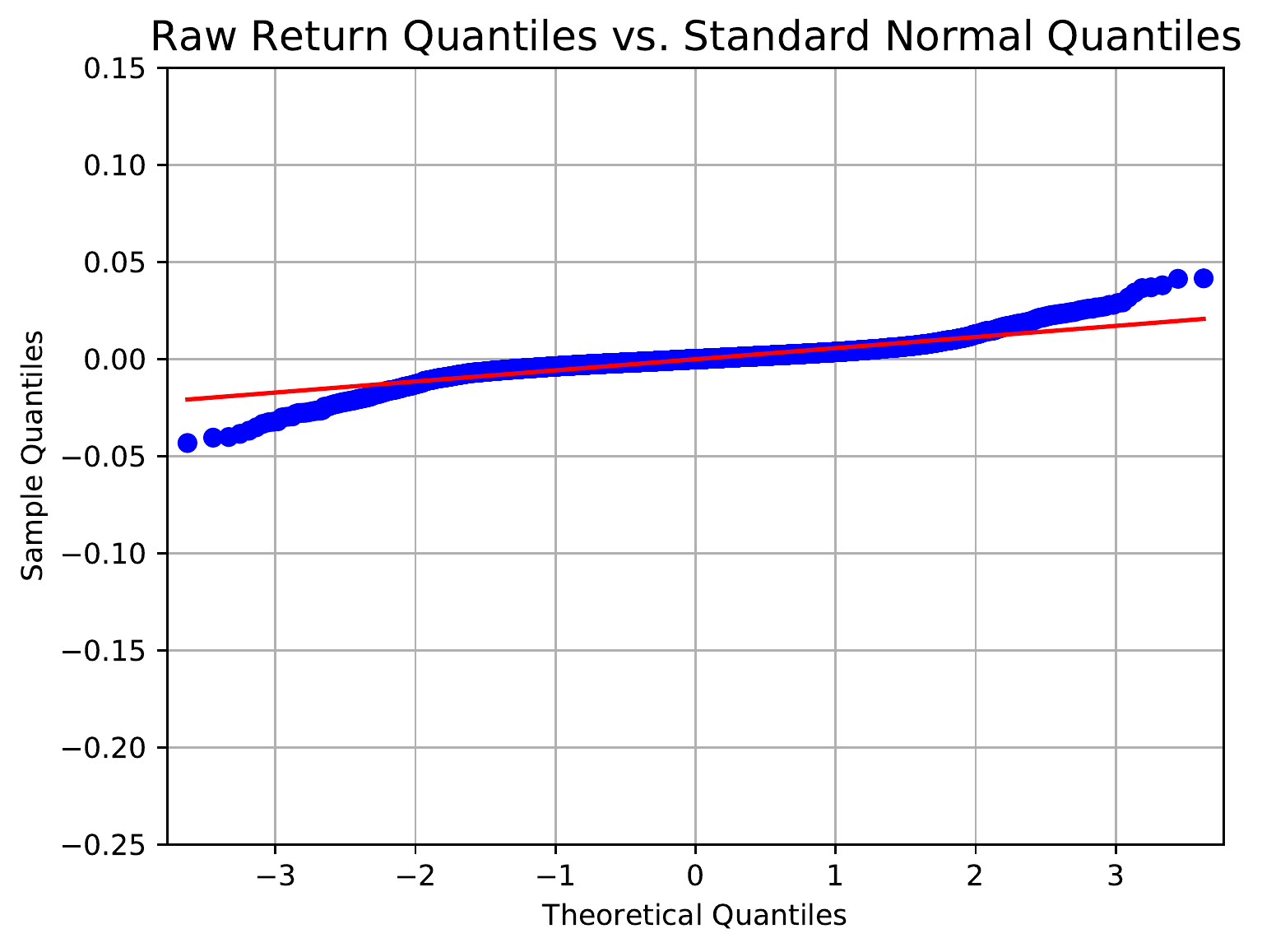}
  \caption{QQ plot of logarithmic returns.}
  \label{fig:fw_dcahpm_qq}
\end{subfigure}
\caption{Macroscopic statistics for a DCA-HPM simulation. Parameters as in table \ref{dcahpm-basic-parameter}.}
\label{fig:fw_auto_qq}
\end{figure}

In this section, we consider the Franke-Westerhoff model as discussed in \citep{franke2012structural}.
Earlier model variants can be found in \citep{franke2009validation, franke2011estimation}. 
The model studies the evolution of two groups of traders, one group with a chartist strategy and the other with a fundamental strategy.
This evolution can be interpreted in the sense of the meta-model introduced in \citep{SABCEMM} as the evolution of two representative agents.
In each time step, a fraction of the traders adapts their investment strategy by a switching process based on socio-economic factors.
These factors are for example a comparison of the agents' estimated wealth or a herding mechanism. 
The logarithmic stock price is then driven by the aggregate excess demand of both groups.
As a distinct feature of the Franke-Westerhoff model, they employ the concept of \textit{structural stochastic volatility} as introduced in \citep{franke2009validation}.
This means that the demand of chartist and fundamentalist has a stochastic component which takes into account heterogeneity within agent groups and uncertain events.
Furthermore, the variance of these random terms differ between both investor groups. \\ \\
The second distinct feature of the Franke-Westerhoff model is the possibility to choose between two well known switching mechanism.
The discrete choice approch (DCA) introduced by Brock and Hommes \citep{brock1997rational} or the transition probability approach (TPA) introduced by Weidlich and Haag \citep{weidlich2012concepts} and Lux \citep{lux1995herd}. \\ \\
The authors have shown in several publications \citep{franke2009validation, franke2011estimation, franke2012structural} that their model fits the statistical features of real financial markets such as volatility clustering or fat-tails of stock returns extremely well.
Furthermore, they have even employed the method of simulated moments \citep{franke2009applying} in order to fit their model parameters to original financial data. \\ \\
First, we present the simulations of the Franke-Westerhoff model with discrete choice approach and the behavioral factors herding (H), predisposition (P) and misalignment (M).
This model choice can be conveniently abbreviated by DCA-HPM.
Figure \ref{fig:fw_auto_qq} shows a plot of the auto-correlation function of raw and absolute logarithmic returns.
The auto-correlation of raw returns is approximately zero which indicates absence of auto-correlation.
In comparison to the raw returns, we obtain a slow algebraic decay in the case of absolute returns.
This is known as volatility clustering. 
In addition, the quantile-quantile plot indicates fat-tails for the logarithmic stock returns.  

\begin{figure}[ht]
\centering
\includegraphics[width=0.49\textwidth]{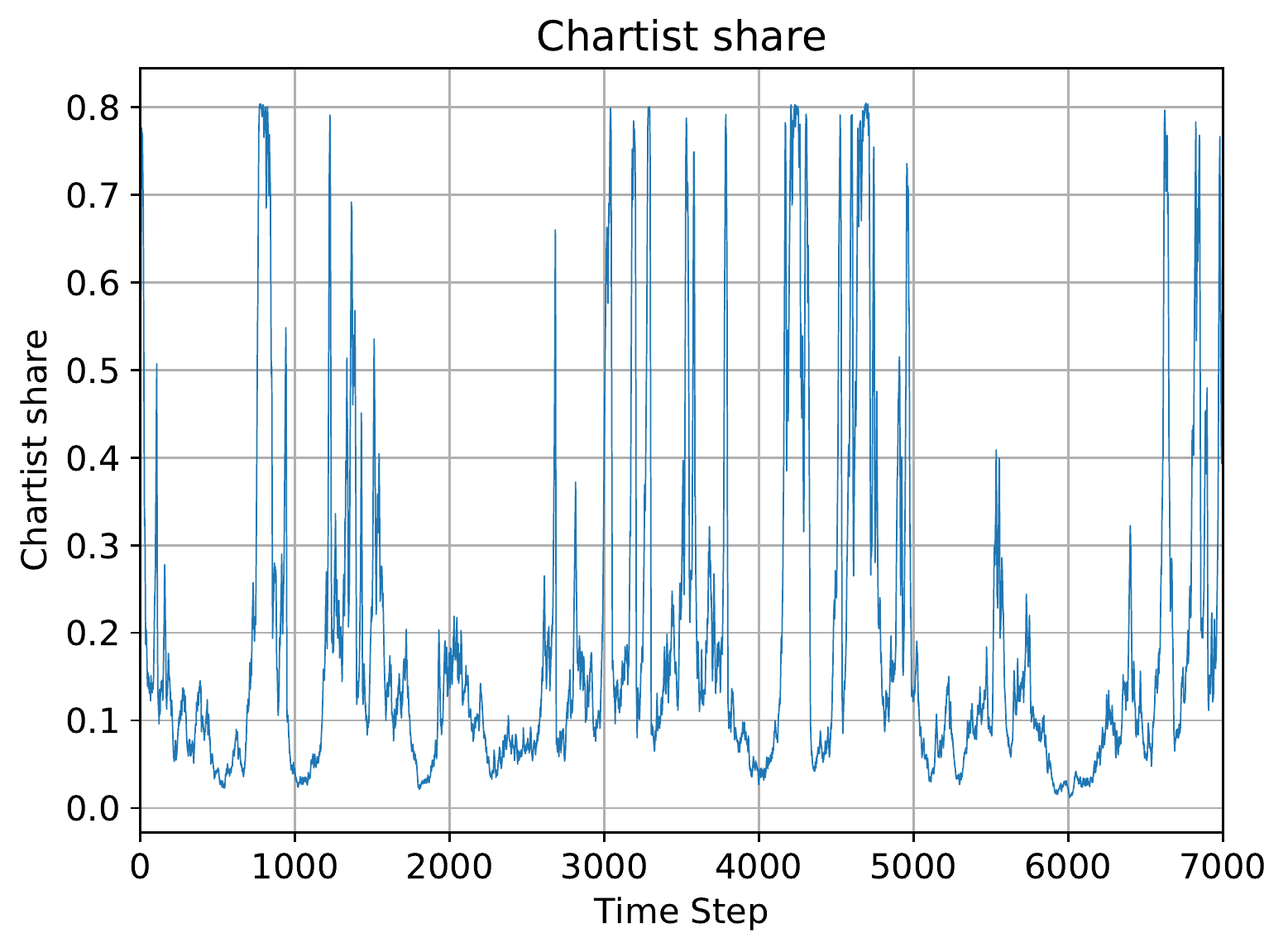}
\includegraphics[width=0.49\textwidth]{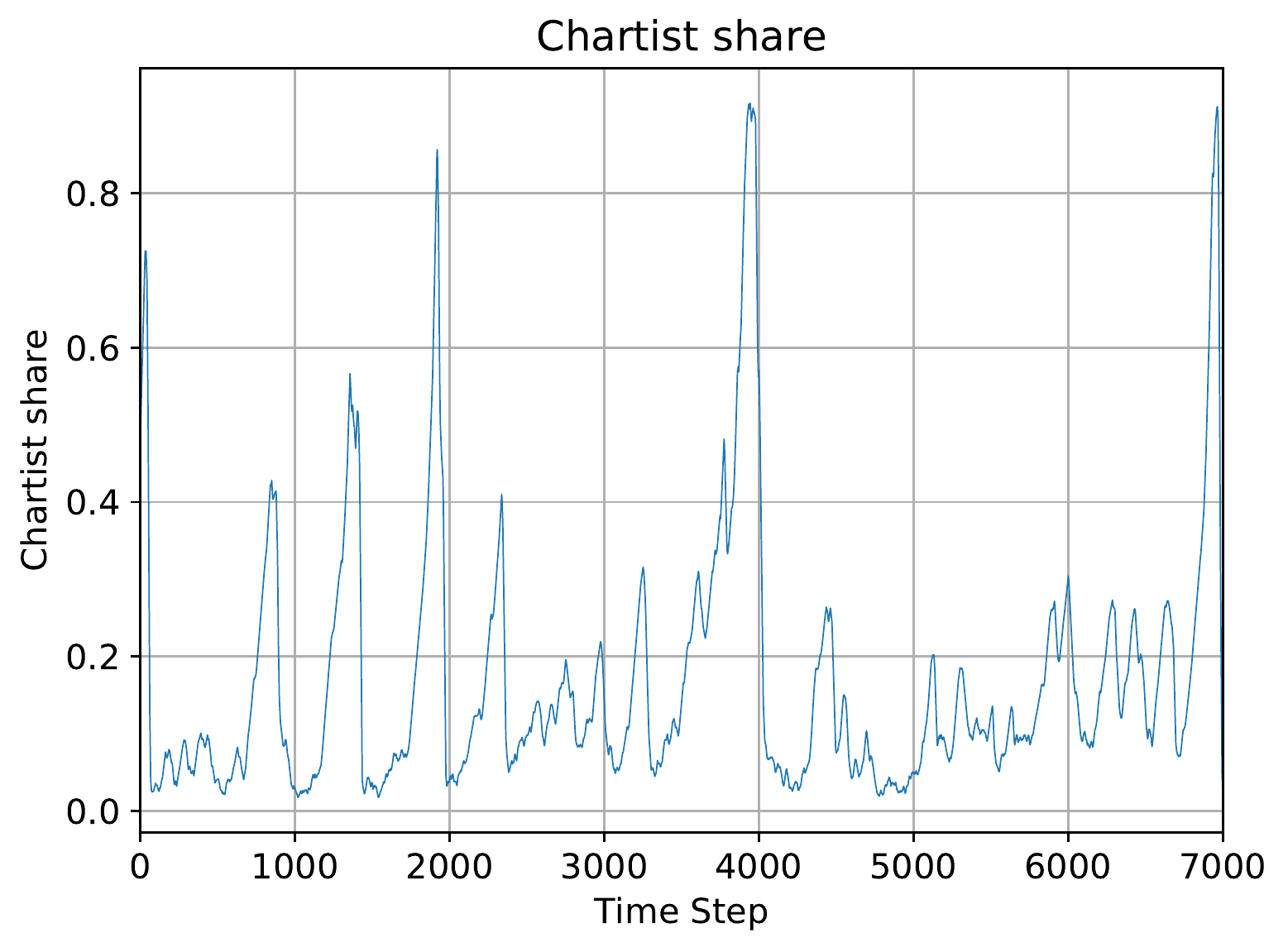}
\caption{Chartist share for DCA-HPM (left hand side) and TPA-HPM (right hand side) switching mechanisms. Parameters as in table \ref{dcahpm-basic-parameter} and \ref{tpahpm-basic-parameter}, respectively.}
\label{fig:dcahpm_tpahpm}
\end{figure}

\begin{figure}[ht]
\centering
\includegraphics[width=0.49\linewidth]{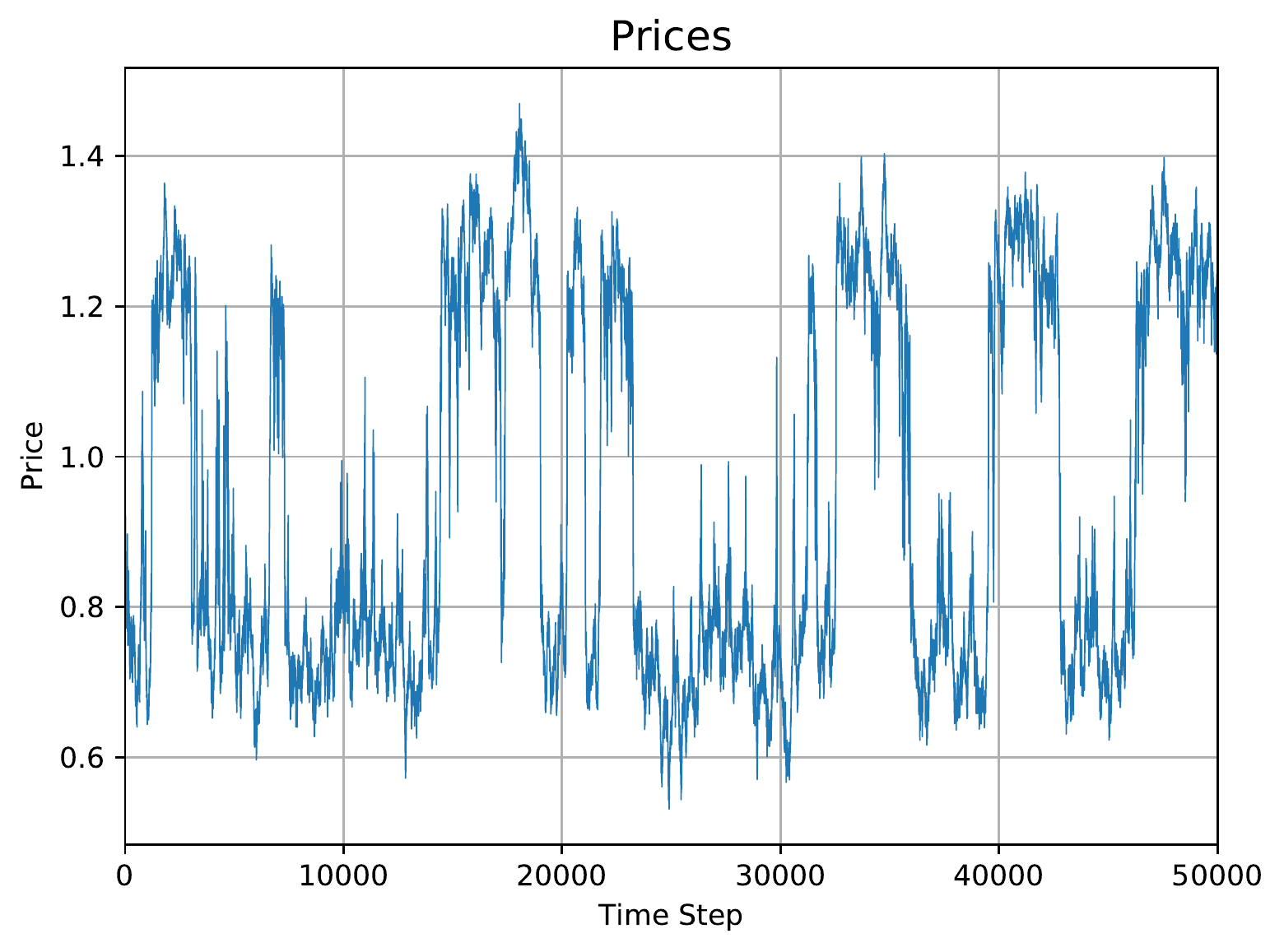}
\caption{Logarithmic prices for a DCA-HPM simulation (50,000 time steps). Parameters as in table \ref{dcahpm-basic-parameter}, with 50,000 time steps.}
\label{fig:fw_longterm}
\end{figure}

As a second aspect, we investigate the impact of both switching mechanisms.
We have simulated the HPM case for the DCA and TPA approach. 
Figure \ref{fig:dcahpm_tpahpm} reveals that the mean fraction of chartists is slightly larger in the DCA approach than in the TPA method (see table \ref{table:fw_mc_results_nolog} as well). 
Furthermore, Figure \ref{fig:dcahpm_tpahpm} shows that the fractions of chartist in the DCA case are much more volatile.
We have used the same random number seed in both simulations.
Qualitatively, the results of both switching mechanisms do not differ.
As presented by Franke and Westerhoff in \citep{franke2011estimation} we run a model contest as well.
We have averaged our simulation over 200 runs.
Table \ref{table:fw_mc_results_nolog} depicts the different values for the excess kurtosis and the hill estimator of logarithmic returns and the average fractions of chartists.
We do not obtain a model which significantly dominates all other choices with respect to reproducing stylized facts.
Finally, we have conducted a test in order to study the long time behavior of the model.
In fact, Figure \ref{fig:fw_longterm} depicts that there is no qualitative change in the evolution of prices in the long run. 


\begin{table}
\begin{center}
\begin{tabular}{l | c c c}
        & excess kurtosis & Hill estimator & avarage chartist share \\ \hline
TPA-W   & {5.9512} & 3.344   &0.2813        \\
TPA-WP  & 7.025           & {3.1957}& 0.2507 \\
TPA-HPM & 8.614           & 2.5833     &   0.1503  \\ \hline
DCA-W   & 8.2023          & {3.173}& 0.2577 \\
DCA-WP  & 7.7600          & {3.1314}& 0.2285\\
DCA-HPM & 10.033          & 2.481   &   0.1674     \\
DCA-WHP & 8.01            & {3.1192}&0.2227 \\ \hline
TPA average    & 7.1967          & 3.041 & 0.2274\\
DCA average     & 7.99          & 2.895 &0.2179 \\ \hline 
\end{tabular}
\end{center}
\caption{Model contest, parameters as in tables \ref{dcaw-basic-parameter} to \ref{tpahpm-basic-parameter}. Each simulation run lasted for 7,000 time steps. The random seed is chosen differently for each repetition, but identical for each of the seven model variants. 
 }
\label{table:fw_mc_results_nolog}
\end{table}

\section{Novel ABCEM Models}\label{NovelModels}
In this section, we demonstrate the flexibility of the SABCEMM simulator by creating new models and adding new features to the Cross model.
More precisely, we consider the Cross agents in combination with new market mechanisms.
We can show that the precise form of the market mechanism can have a direct impact on the appearance of stylized facts. 
In addition, we modify the Cross agents by adding a wealth evolution to each agent and study the statistical properties of the wealth evolution.

\paragraph{Wealth Evolution}
\label{NM-Cross-WE}

\begin{figure}[h!]
    \begin{center}
        \includegraphics[width=0.7\textwidth]{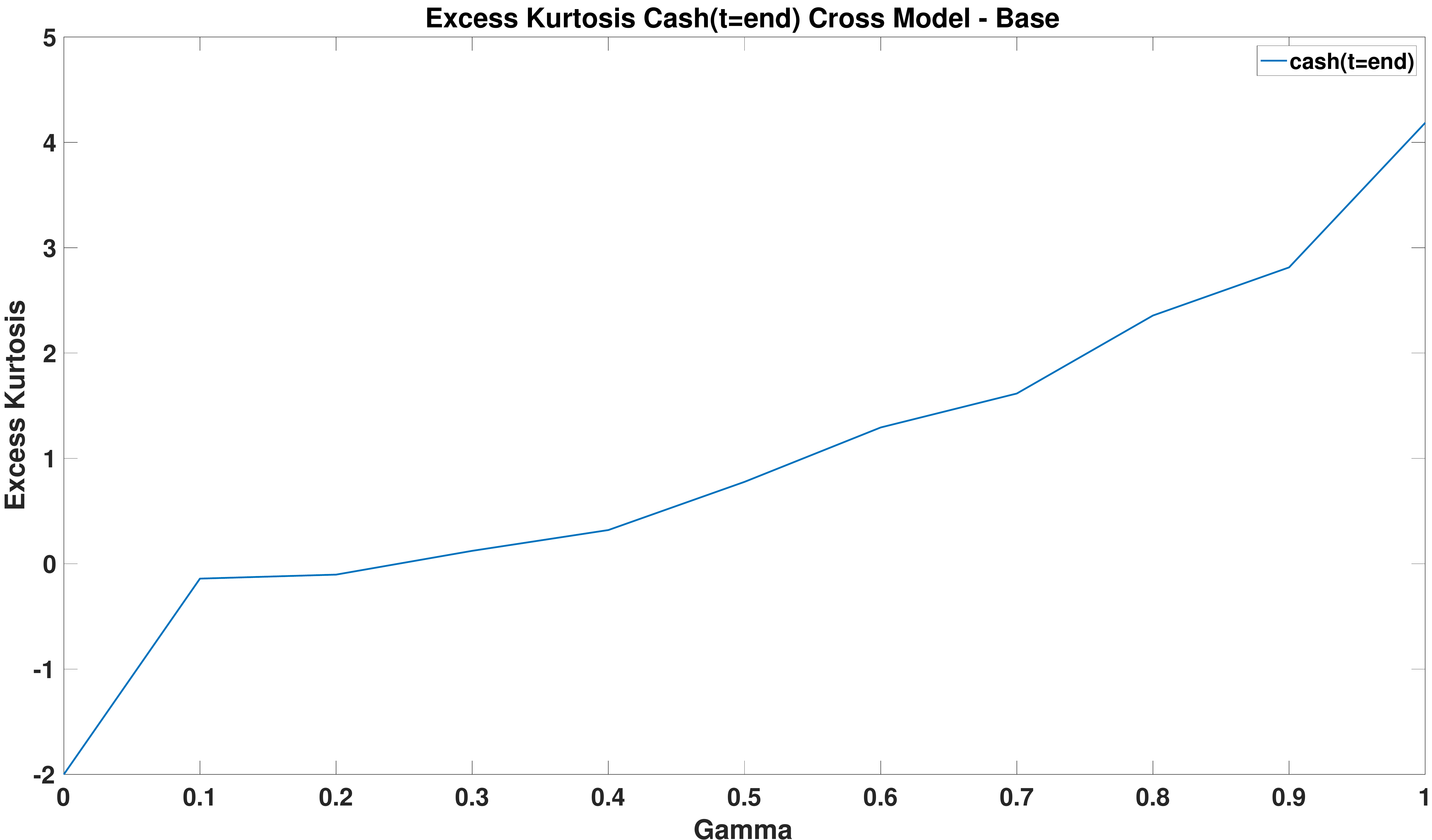}
        \caption{Excess kurtosis for the wealth at the final timestep of the Cross model. Parameters as in \cref{cross-basic-parameter} with $w_i(t=0)=1$ and $r = 0.01$.}
        \label{NM-Cross-WE-base}
    \end{center}
\end{figure}

\begin{figure}[h!]
    \begin{center}
        \includegraphics[width=0.7\textwidth]{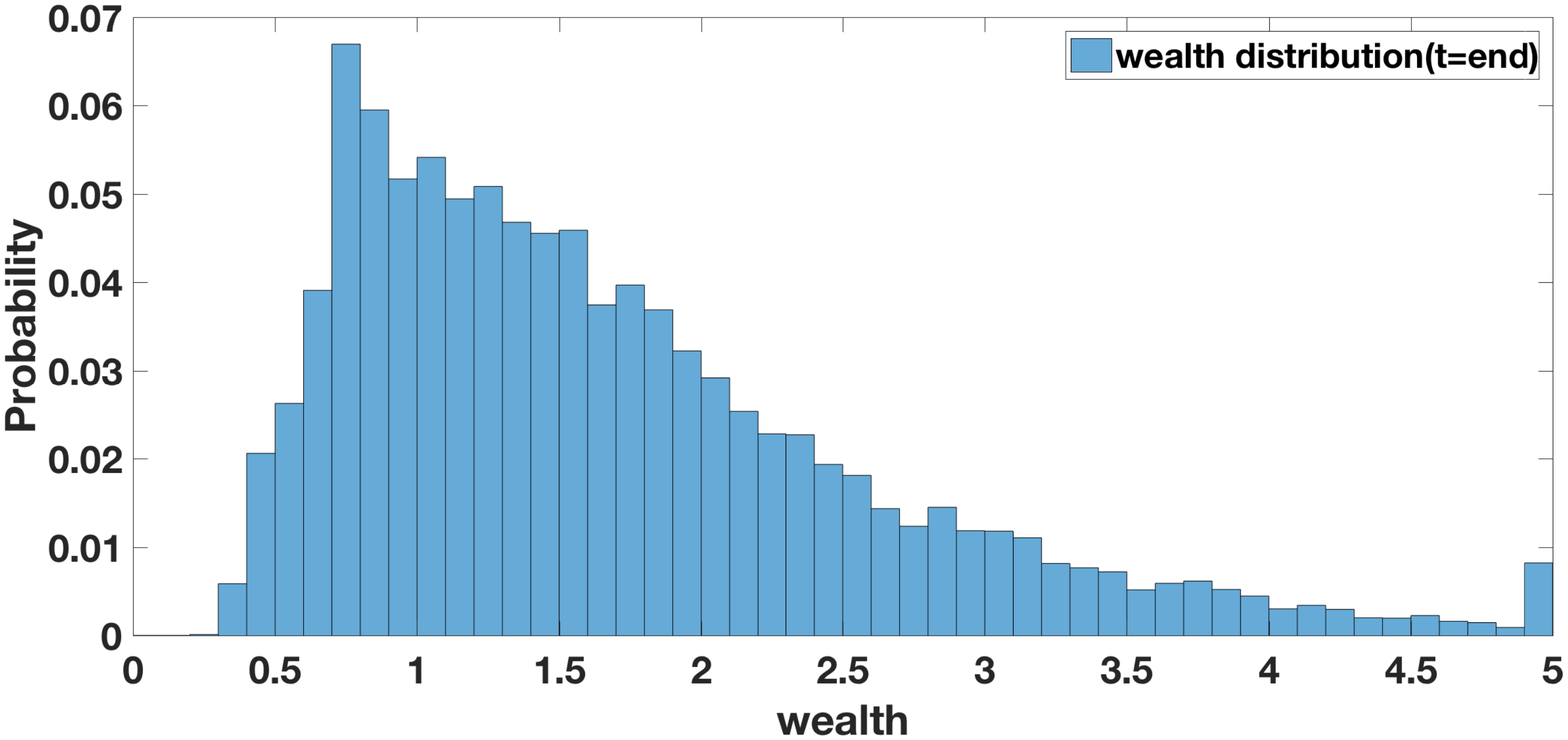}
        \caption{Histogram of the wealth distribution at the final timestep. Parameters as in \cref{cross-basic-parameter} with $w_i(t=0)=1$ and $r = 0.01$.}
        \label{NM-Cross-WE-histo}
    \end{center}
\end{figure}

\begin{figure}[h!]
    \begin{center}
        \includegraphics[width=0.7\textwidth]{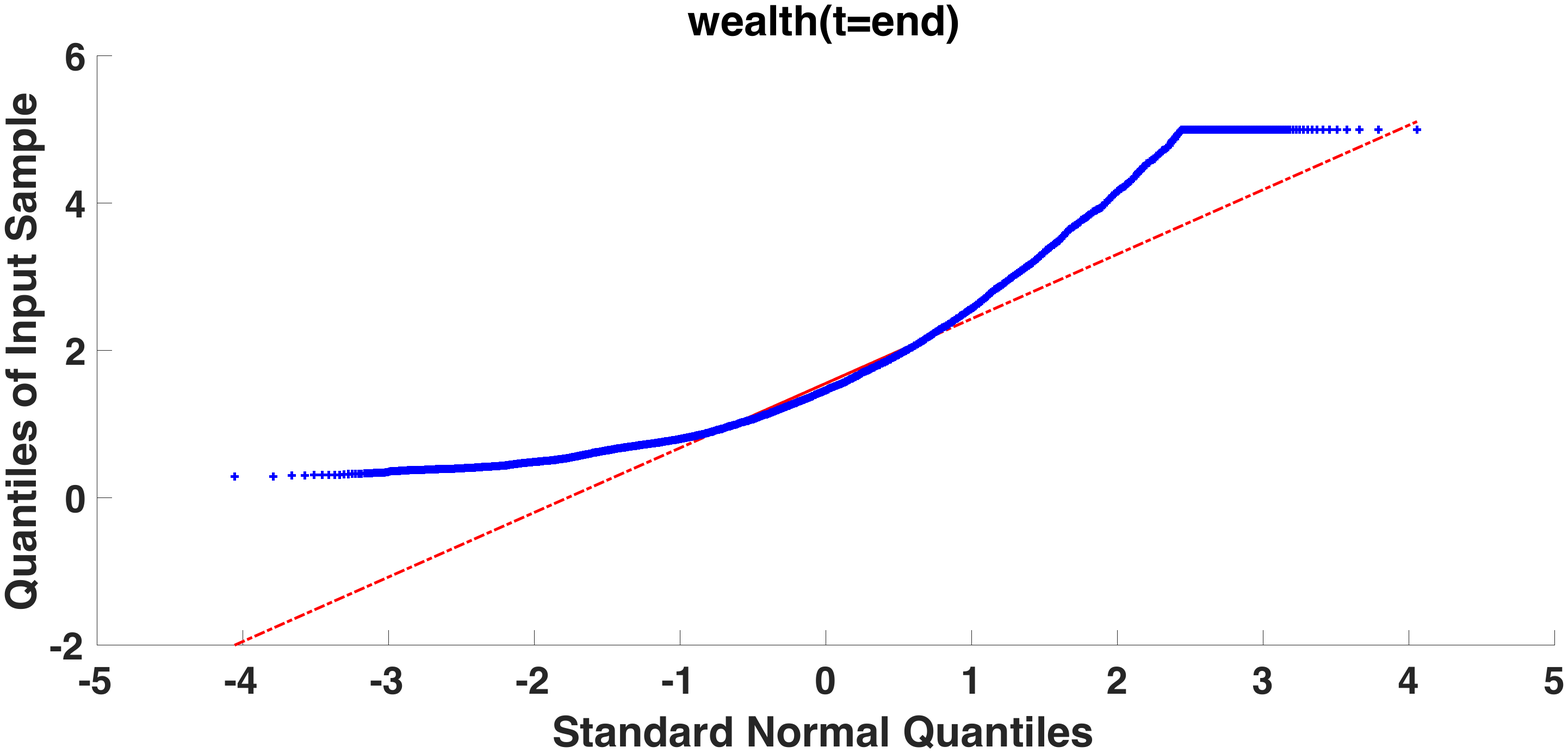}
        \caption{QQ-Plot of the wealth distribution. Parameters as in \cref{cross-basic-parameter} with $w_i(t=0)=1$ and $r = 0.01$. }
        \label{NM-Cross-WE-qq}
    \end{center}
\end{figure}

We consider the original Cross model as defined in appendix \ref{appendixModel}.
We add a wealth evolution of the type 
$$
w_{i}(t_{k+1}) = w_i(t_k)+ \Delta t\ \left[(1-\gamma)\ r + \gamma \frac{S(t_k)-S(t_{k-1})}{\Delta t\ S(t_k)}\right] w_i(t_k),
$$
to each Cross agent $i=1,...,N$.
The positive constant $r>0$ denotes the interest rate and $\gamma\in (0,1)$ a fixed fraction of stock investments for all agents.
The goal is to study the influence of the non-Gaussian return distribution on the wealth distribution.
\changes{We have plotted the empirical wealth distribution in Figure \ref{NM-Cross-WE-histo}.}
With increasing $\gamma$, we obtain an increasing excess kurtosis of the wealth distribution.
In fact, the excess kurtosis approaches the excess kurtosis of the stock price, which is approximately 6.
The results averaged over 100 runs are shown in Figure \ref{NM-Cross-WE-base}.
\changes{
The qq-plot in Figure \ref{NM-Cross-WE-qq} clearly shows the asymmetrically fat-tail behavior of the wealth distribution for a fixed $\gamma=1$.
Thus, the tails of the wealth distributions are fat for positive values and slim for negative values.
This behavior reveals that the tail behavior is translated from the price distribution to the wealth distribution and heavily depends on the investment fraction $\gamma$.
}

For future research, we could generalize the microscopic excess demand of the Cross agents by adding a wealth dependency.
This would lead to an additional coupling between wealth and stock price behavior. 
\clearpage

\begin{figure}[h!]
    \begin{center}
        \includegraphics[width=0.7\textwidth]{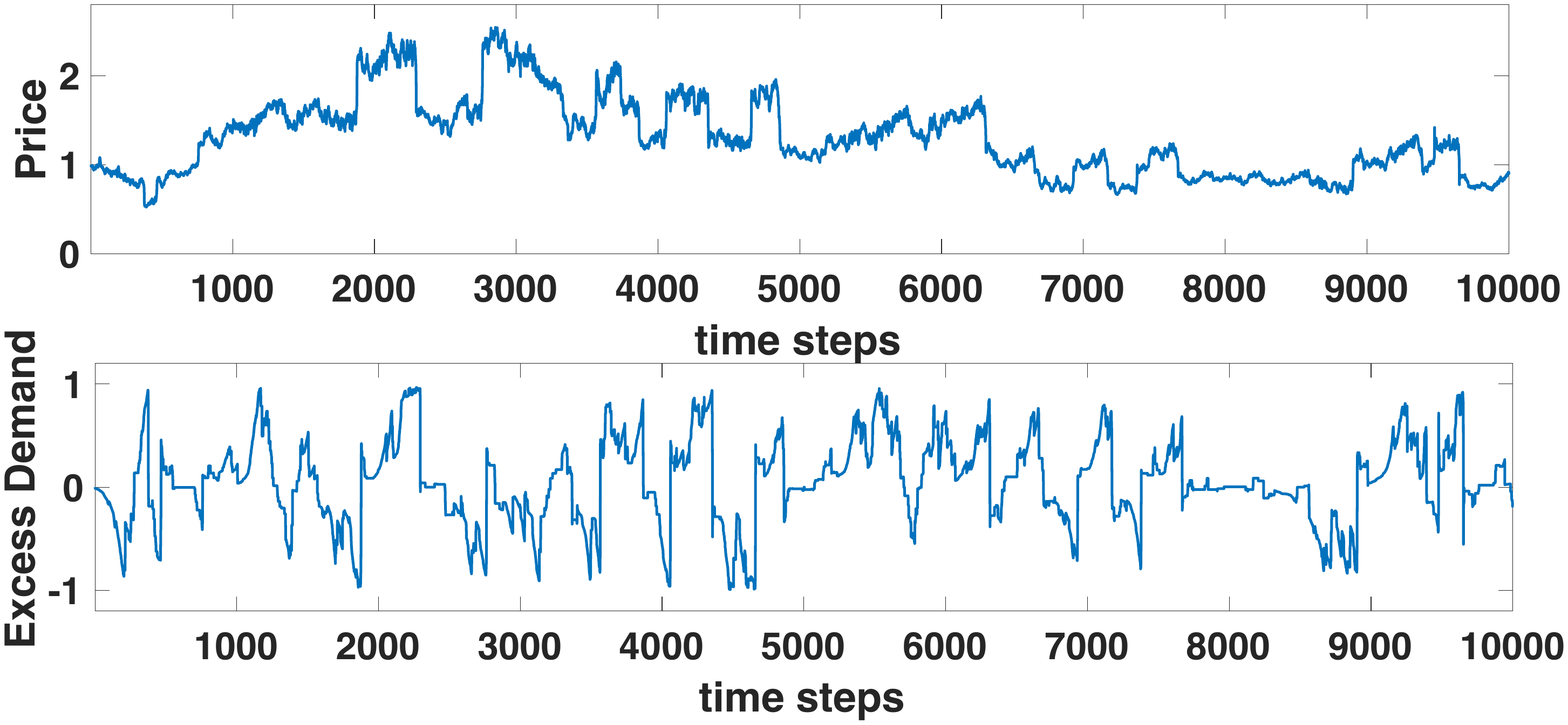}
        \caption{Cross agents with SDE pricing rule. Parameters as in \cref{cross-basic-parameter} except $\theta = 2$.}
        \label{NM-Cross-SDE-Hetero}
    \end{center}
\end{figure}

\begin{figure}[h!]
    \begin{center}
        \includegraphics[width=0.7\textwidth]{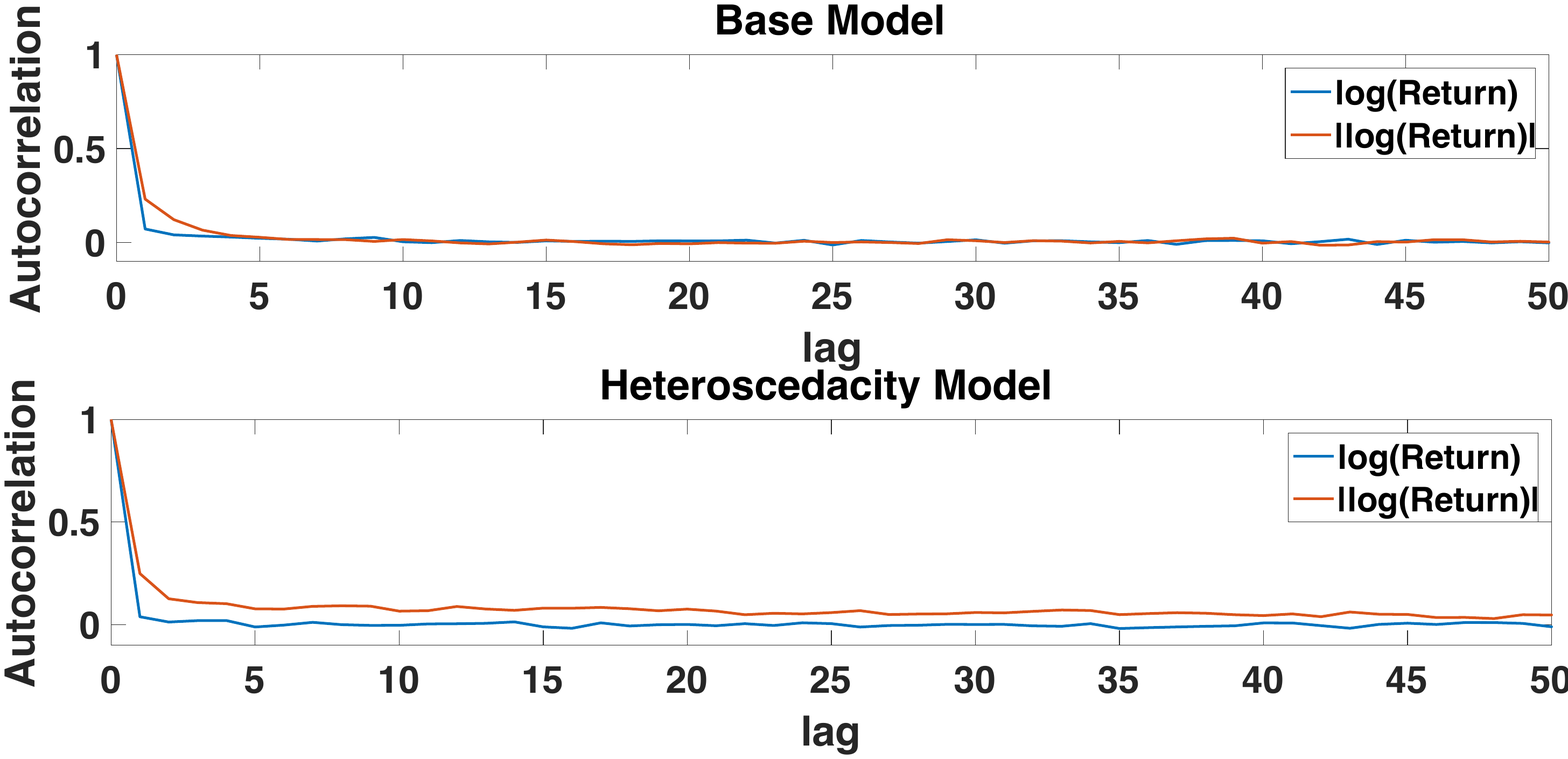}
        \caption{Auto-correlation of log-returns and absolute log-returns in the base model (upper graph, parameters see \cref{cross-basic-parameter}) and full model (parameters see \cref{cross-basic-parameter} except $\theta =2$). We obtain the same behavior as previously in section \ref{OM-Cross}.}
        \label{NM-Cross-SDE-AutoCorr}
    \end{center}
\end{figure}

\paragraph{SDE Discretization}
In this section, we again consider the Cross agents, but we change the clearance mechanism. 
We set the pricing rule to be
$$
S(t_{k+1}) = S(t_k)+ \Delta t\   F_{Cross}(t_k,S, ED) + \sqrt{\Delta t}\  S(t_k)\  (1+ \theta\ | ED(t_k) |)\ \eta,\quad \eta\sim \mathcal{N}(0,1). 
$$
We choose the Euler-Maruyama discretization of the SDE:
$$
dS = F_{Cross}(S,ED)\ dt + S\ (1+\theta\ | ED|)\ dW,
$$
where $W$ is a Wiener process and the SDE should be interpreted in the It\^{o} sense.
For our first simulation, we set the drift operator $F$ to read:
$$
F_{Cross}^1(t,S,ED) := S(t)\ \frac{d}{dt} ED(t).
$$
As the Figures \ref{NM-Cross-SDE-Hetero} and \ref{NM-Cross-SDE-AutoCorr} reveal, the behavior of this model is identical to the original Cross model. 
Thus, we have obtained a pricing rule which is a proper time discretization of a time continuous model. 

In a second test we set the drift coefficient to
$$
F^2_{Cross}(t,S, ED) := S(t)\  ED(t).
$$
We study the different impact of the two models on the stock price behavior. 
Our studies reveal that the the choice $F^2_{Cross}$ leads to Gaussian price behavior.
We measured this using the excess kurtosis, which we averaged over $100$ runs (see table \ref{tabEK}). 
This is an interesting result, since it reveals the great influence of the drift coefficient on the stock price behavior. 

\begin{table}[h!]
\begin{center}
    \begin{tabular}{ | l | c |}
        \hline
          & excess kurtosis \\ \hline
        $F^1_{Cross},\ \theta=0$ & 27.0434  \\ \hline
        $F^1_{Cross},\ \theta=2$ & 22.5530 \\ \hline
        $F^2_{Cross},\ \theta=0$ & -0.0044 \\ \hline
        $F^2_{Cross},\ \theta=2$ & 1.2580  \\ \hline
    \end{tabular}
 \caption{Averaged excess kurtosis over 100 runs with different drift functions. }\label{tabEK}
\end{center}
\end{table}

%% file: conclusion.tex
\section{Conclusion}
\label{sec-conclusion}
\changes{
We have presented various simulation results using the recently introduced simulator SABCEMM.
We have provided a brief introduction to the SABCEMM software. Afterwards, we have carried out several simulations of the 
LLS model, Cross model and Franke-Westerhoff model. In order to verify that the Cross model is not prone to finite size effects, we ran simulations of up to several million agents. We utilized the well-separated, modular design of SABCEMM, allowing for the recombination of building blocks, to create novel ABCEM models by simply interchanging the market mechanism.\\ \\
For the previously published ABCEM models, our results and the obtained stylized facts coincide with the findings in literature.
The extended numerical studies such as simulations with many agents or long time simulations have supported the validity of these well-known models.\\
The study of the new ABCEM models based on the agent design of the Cross model has revealed several novel insights.
Our studies have shown that the fat-tail property of the stock return influences the tail of the wealth distribution.}
Interestingly, the fraction of investments in the stock return determines the size of the tail. 
Hence, the more money is invested in stocks the more prominent is the tail in the wealth distribution. 
Furthermore, we have seen that the choice of the market mechanism heavily influences the fat-tail property of the stock return data.
We conclude that the precise form of the market mechanism is of eminent importance in order to generate realistic stock price data. \\ \\
These observations lead us to the conclusions that the combination of further building blocks may help to find the origins of stylized facts and to understand the mechanism which create them.

%% file: appendix.tex
\section{Appendix}
\label{sec-appendix}

\subsection{Mathematical Appendix}

\begin{definition}\label{def:cor}
The auto-correlation for the stationary stochastic process $R(t),\ t>0$ is defined by:
\begin{align*}
C(l):= Corr(R(t+l),R(t))=\frac{Cov(R(t+l), R(t))}{E[(R(t)-\bar{R})^2]}= \frac{E[(R(t+l)-\bar{R})\ (R(t)-\bar{R})]}{E[(R(t)-\bar{R})^2]},\ l>0
\end{align*}
The correlation $Corr$ is given by the normalized covariance $Cov$ of two random variables. The auto-correlation function $C(l)\in[-1,1]$ depends on the time shift called lag $l>0$ of the stochastic process.
\end{definition}

\begin{definition}\label{def:exkurt}
The excess kurtosis is defined as the normalized fourth moment of the stationary stochastic stock returns process $R$ minus a correction term, defined by
\begin{align*}
&\kappa:= \frac{E[(R-\bar{R})^2]}{\sigma^4}-3,\\
& \bar{R}:= E[R],\\
&\sigma^2:= E[(R-\bar{R})^2]. 
\end{align*}
\end{definition}

\begin{definition}
\label{def:hill_estimator}
The Hill estimator of a  sample $X\in \R^n_{>0}$ sorted as $x_1 \geq x_2 \geq ... \ x_n$ is defined as 
$$ H := \left(\frac{1}{k}\sum_{i=1}^{k}\ln(x_i) - \ln( x_k)\right)^{-1},  $$
with $k:=\lfloor 0.05*n\rfloor$. For details we refer to \citep{hill1975simple}. 
\end{definition}

\subsection{Models}
\label{appendixModel}

\paragraph{Cross Model}
We present the Cross model as defined in \citep{cross2005threshold}.\\ \\
We assume a fixed number of $N\in\N$ agents. Each agent decides in each time step, whether he wants to be long or short in the market. Thus, the investment propensity $\sigma_i,\ 1\leq i\leq N$ of each agent switches between $\sigma_i=\pm 1$. The excess demand of all investors at time $[0,\infty)$ is then defined as:
\[
ED(t_k):= \frac{1}{N}\sum\limits_{i=1}^N\sigma_i(t_k).
\] 
Furthermore, the model introduces two pressures, the \textit{herding pressure} and the \textit{inaction pressure}, which control the switching mechanism.\\
The \textit{inaction pressure} is defined by the interval
\[
I_i=\left[ \frac{m_i}{1+\alpha_i}, m_i\ (1+\alpha_i)\right],
\] 
where $m_i$ denotes the stock price of the last switch of agent $i$ and $\alpha_i>0$ is the so called \textit{inaction threshold}.
The \textit{herding pressure} is given by:
\[ \begin{cases} c_i(t_{k+1})= c_i(t_k)+ \Delta t |ED(t_k)|,& \text{if}\ \sigma_i(t_k)\ ED(t_k)<0\\
			c_i(t_{k+1})=c_i(t_k),& \text{otherwise}.
			 \end{cases}
			\]
In addition, one defines the  \textit{herding threshold} $\beta_i$. The thresholds are chosen once randomly from an i.i.d. random variable, which is uniformly distributed. 
\begin{align*}
\alpha_i\sim \Unifc(A_1,A_2), \ A_2>A_1>0,\\
\beta_i\sim \Unifc(B_1,B_2), \ B_2>B_1>0.
\end{align*}

The constants $B_1$ and $B_2$ have to scale with time, since they correspond to the time units an investor can resist the herding pressure.
\begin{align*}
& B_1:= b_1\cdot \Delta t,\\
&B_2:= b_2\cdot \Delta t.
\end{align*}
\paragraph{Switching mechanism}
	The switching is then induced if 
	\[
	c_i>\beta_i\ \text{or}\  S(t)\notin I_i.
	\]
After a switch the herding pressure is reset to zero and the inaction interval gets updated as well. The stock price is then driven by the excess demand:
\begin{align*}
&S(t_{k+1})=S(t_k)\ \exp\{  (1+\theta\ |ED(t_k)| )\ \sqrt{\Delta t}\ \eta(t_k)+\kappa\ \Delta t\ \frac{\Delta ED(t_k)}{\Delta t} \},\\
&\sqrt{\Delta t}\ \eta\sim \mathcal{N}(0,\Delta t)\\
&\Delta ED(t_k):=\frac{1}{N}\sum\limits_{i=1}^N \sigma_i(t_k)-\frac{1}{N}\sum\limits_{i=1}^N \sigma_i(t_{k-1}),
\end{align*}
where $\kappa$  denotes the market depth and $\Delta t >0$ the time step.\\

\paragraph{Cross model extensions:}
One alternative pricing function is given by:
\begin{align*}
S(t_{k+1})=S(t_k)\ + \Delta t\  \kappa\ \frac{\Delta ED(t_k)}{\Delta t} \ S(t_k)+ \sqrt{\Delta t}\ F_{Cross}(t_k,S,ED)\ S(t_k)\  \eta,\\
\end{align*}
Furthermore, we have added the wealth evolution, for a fixed interest rate $r>0$ and fixed investment fraction $\gamma\in (0,1)$:
$$
w_{i}(t_{k+1}) = w_i(t_k)+ \Delta t\ \left[(1-\gamma)\ r + \gamma \frac{S(t_k)-S(t_{k-1} )}{\Delta t\ S(t_k)}\right] w_i(t_k).
$$

\paragraph{LLS Model}
We have implemented the model as defined in \citep{levy1994microscopic, levy1995microscopic}. 
We have added one possible time scale to the model. In order to obtain the original model one needs to set $\Delta t=1$.\\ \\
The model considers $N\in\N$ financial agents who can invest $\gamma_i\in [0.01, 0.99],\ i=1,...,N$ of their wealth $w_i\in\R_{>0}$ in a stocks and have to invest $1-\gamma_i$ of their wealth in a safe bond with interest rate $r\in(0,1)$. The investment propensities $\gamma_i$ are determined by a utility maximization and the wealth dynamic of each agent at time $t\in[0,\infty )$ is given by
\begin{footnotesize}
\begin{align*}
{w}_i(t_k)&=w_i(t_{k-1})\\
\quad &+\Delta t \left((1-\gamma_i(t_{k-1}))\ r\ w_i(t_{k-1})+\gamma_i(t_{k-1})\ w_i(t_{k-1})\ \underbrace{\frac{\frac{{S}(t_k)-S(t_{k-1})}{\Delta t}+D(t_k)}{S(t_k)}}_{=:x(S,t_k,D)}\right).
\end{align*} 
 \end{footnotesize}
The dynamics is driven by a multiplicative dividend process. Given by:
\[
D(t_k):=(1+\Delta t\  \tilde{z})\  D(t_{k-1}),
\]
 where $\tilde{z}$ is a uniformly distributed random variable with support $[z_1,z_2]$.  
The price is fixed by the so called \textit{market clearance condition}, where $n\in\N$ is the fixed number of stocks and $n_i(t)$ the number of stocks of each agent.  
\begin{align}
n=\sum\limits_{i=1}^N n_i(t_k)=\sum\limits_{k=1}^N \frac{\gamma_k(t_k)\ w_k(t_k)}{S(t_k)}. \label{fixedpointLLS}
\end{align}
The utility maximization is given by
\[
\max\limits_{\gamma_i \in [0.01,0.99]} E[\log(w(t+\Delta t,\gamma_i,S^h))].
\]
with
\begin{align*}
E[\log(w(t_{k-1}, \gamma_i,S^h))]=\frac{1}{m_i} \sum\limits_{j=1}^{m_i} U_i\Bigg(&(1-\gamma_{i}(t_k)) w_{i}(t_k,S^h) \left(1+r\Delta t\right)\\ 
&+\gamma_{i}(t_k) w_{i}(t_k, S^h) \Big(1+x\big(S,t_k-j\Delta t,D\big)\ \Delta t\Big)\Bigg).
\end{align*}

The constant $m_i$ denotes the number of time steps each agent looks back. Thus, the number of time steps $m_i$ and the length of the time step $\Delta t$ defines the time period each agent extrapolates the past values. The superscript $h$ indicates, that the stock price is uncertain and needs to be fixed by the market clearance condition.
Finally, the computed optimal investment proportion gets blurred by a noise term.
\[
\gamma_i (t_k)=\gamma_i^{*}(t_k)+\epsilon_i,
\]
where $\epsilon_i$ is distributed like a truncated normally distributed random variable with standard deviation $\sigma_{\gamma}$.

\paragraph{Utility maximization}
Thanks to the simple utility function and linear dynamics we can compute the optimal investment proportion in the cases where the maximum is reached at the boundaries.
The first order necessary condition is given by:
$$
f(\gamma_i):= \frac{d}{dt} E[\log(w(t_{k+1}, \gamma_i,S^h))] = \frac{1}{m_i} \sum\limits_{j=1}^{m_i} \frac{\Delta t\ (x\big(S,t_k-j\Delta t,D\big)-r)}{\Delta t\ (x\big(S,t_k-j\Delta t,D\big)-r)\ \gamma_i+ 1+\Delta t\ r}.
$$
 Thus, for $f(0.01)<0$ we can conclude that $\gamma_i=0.01$ holds. In the same manner, we get $\gamma_i=0.99$, if $f(0.01)>0$ and $f(0.99)>0$ holds. 
 Hence, solutions in the interior of $[0.01, 0.99]$ can be only expected in the case: $f(0.01)>0$ and  $f(0.99)<0$. This coincides with the observations in \citep{samanidou2007agent}.

\paragraph{Franke-Westerhoff model}
The Franke-Westerhoff model \citep{franke2011estimation} considers tow types of agents, chartists and fundamentalists. The demand of each agent reads
\begin{align}
d^f (t_k)= \phi (P_f-P(t_k)) + \epsilon_k^f, \quad \phi \in \mathbb{R}^+, \quad \epsilon_k^f \sim \mathcal{N}(0, \sigma_f^2), \\
d^c (t_k)= \chi (P(t_k) - P(t_{k-1})) + \epsilon_k^c, \quad \chi \in \mathbb{R}^+, \quad \epsilon_k^c \sim \mathcal{N}(0, \sigma_c^2),
\end{align}
where $P(t_k)$ denotes the logarithmic market price and $P_f$ denotes the fundamental price. The noise terms $\epsilon_k^f$ and $\epsilon_k^c$ are normally distributed, with zero mean and different standard deviations $\sigma_c^2$ and $\sigma_f^2$. The second important features are the fractions of the chartist or fundamental population. In that sense the two agents can be seen as representative agents of a population. The fraction of chartists $n^C(t_k)\in[0,1]$ and the fraction of fundamentalitst $n^F(t_k)\in[0,1]$ have to fulfill $n^C(t_k)+n^F(t_k)=1$. 
Hence, the excess demand can be define as:
\begin{equation}
ED^F(t_k) := \frac12 (ed^f(t_k) + ed^c(t_k))
\end{equation}
\begin{align}
ed^f(t_k) := 2\ n^f(t_k) d^f(t_k) \nonumber \\
ed^c(t_k) := 2\ n^c(t_k d^c(t_k). 
\end{align}
The pricing equation is then given by the simple rule
\begin{equation}
P(t_k) = P(t_{k-1}) + \mu ED^F(t_{k-1}).
\end{equation}
Finally, we need to specify the switching mechanism. We have implemented two possible switching mechanisms, the transition probability approach (TPA)  \citep{weidlich2012concepts, lux1995herd}
and the discrete choice approach (DCA) \citep{brock1997rational} approach. In both cases we consider the so called switching index $a(t_k)\in\R$ which describes the attractiveness of the fundamental strategy over the chartist strategy. Thus, a positive $a(t_k)$ reflects an advantage of the fundamental strategy in comparison to the chartist and if $a(t_k)$ is negative we have the opposite situation. 
\paragraph{DCA}
In the DCA case we obtain
\begin{align}
	n^f(t_k) &= \frac{1}{1+exp(-\beta a(t_{k-1}))}, \nonumber  \\
	n^c(t_k) &= \frac{1}{1+exp(\beta a(t_{k-1}))},
\end{align}
where the parameter $\beta>0$ measures the intensity of choice. 
\paragraph{TPA}
In the TPA case, we first define switching probabilities
\begin{equation}
\begin{aligned}
\pi^{cf}(a(t_{k-1})) &= min[1, \nu  \exp(a(t_{k-1}))],  \nonumber \\
\pi^{fc}(a(t_{k-1}) &= min[1, \nu  \exp(- a(t_{k-1}))], \nonumber
\end{aligned}
\end{equation}
where $\pi^{xy}$ is the probability that an agent with strategy x switches to strategy y. The flexibility parameter $\nu$ is a scaling factor for $a(t_k)$. 
Then the time evolution of chartist and fundamentalist shares is given by:
\begin{equation}
\begin{aligned}
n^f(t_k) &= n^f(t_{k-1}) + n^c(t_{k-1}) \pi^{cf}(a(t_{k-1})) - n^f(t_{k-1}) \pi^{fc}(a(t_{k-1})) \nonumber \\
n^c(t_k) &= n^c(t_{k-1}) + n^f(t_{k-1}) \pi^{fc}(a(t_{k-1})) - n^c(t_{k-1}) \pi^{cf}(a(t_{k-1}))
\end{aligned}
\end{equation}
Finally, we have to specify how the switching index $a(t_k)$ is calculated. The switching index $a(t_k)$, encodes how favourable a fundamentalist strategy is over a chartist strategy. 
The switching index is determined linearly out of the the four principles \textit{wealth comparison,\ predisposition, herding and misalignment}. 
$$
\alpha(t_k) =\alpha_w\ (w^f(t_k)-w^c(t_k))+\alpha_0+ \alpha_h\ (n^f(t_k)-n^c(t_k))+ \alpha_m\ (P(t_k)-P_f)^2,
$$
where $\alpha_w,\alpha_0, \alpha_h,\alpha_m>0$ are weights respectively scaling factors. The sign  $\alpha_p\in\R$ determines the predisposition with respect to a fundamental or chartist strategy. 
The hypothetical wealth $w^f(t_k), w^c(t_k)$ is determined as follows:
\begin{align*}
& w^f(t_k) :=  m\ w^f(t_{k-1}) + (1-m) g^f(t_k),\\
& g^f(t_k):= [exp(P(t_k))-exp(P(t_{k-1}))] d^f(t_{k-2}),\\
& w^c(t_k) :=  m\ w^c(t_{k-1}) + (1-m) g^c(t_k),\\
& g^c(t_k):= [exp(P(t_k))-exp(P(t_{k-1}))] d^c(t_{k-2}).\\
\end{align*}
Here, the memory variable $m\in[0,1]$ weights the past performance with the most recent return. 
For details regarding the modeling we refer to \citep{franke2011estimation}.

\subsection{Parameter sets}
\paragraph{Cross Model}

\begin{table}[h!]
	\begin{subtable}[b]{0.45\textwidth}
	    \begin{center}
	        \begin{tabular}{|c||c|}
	            \hline
	            Parameter & Value\\
	            \hline
	            \hline
	            $N$ & $1000$ \\
	            \hline
	            $A_1$& $0.1$\\
	            \hline 
	            $A_2$ & $0.3$\\
	            \hline
	            $b_1$& $25$\\
	            \hline 
	            $b_2$ & $100$\\
	            \hline 
	            $w_i(t=0)$ & $1 \quad \forall 1\leq i\leq N$\\
	            \hline 
	            time steps & $10,000$ \\
	            \hline
	            $\Delta t$ &$ 4\cdot 10^{-5} $\\
	            \hline
	            $\kappa$ & $0.2$\\
	            \hline
	            $\theta$& $0$\\
	            \hline
	            $S(t=0)$ & $1$\\
	            \hline 
	        \end{tabular} 
		\end{center}
		\caption{Parameters of Cross model. }
	\end{subtable}
	\hspace{0.2cm}
	\begin{subtable}[b]{0.45\textwidth}
		\begin{center}
	        \begin{tabular}{|c||c|}
	            \hline
	            Variable & Initial Value\\
	            \hline
	            \hline
	            $ED(t=0)$& $\frac{1}{N}\sum\limits_{i=1}^N \gamma_i(0)$\\
	            \hline
	            $c_i(t=0)$ & $B_1+\texttt{rand}\ (B_2-B_1),\ \forall 1\leq i\leq N$\\
	            \hline
	            $m_i(t=0)$ & $S(t=0),\ \forall 1\leq i\leq N$\\
	            \hline
	            $\sigma_i(t=0)$ & $ \Unifd(\{-1,1\})$ \\
	            \hline
	        \end{tabular}
	    \end{center}
	    \caption{Initial values of Cross model.}
	\end{subtable}
	\caption{Cross basic setting.}
    \label{cross-basic-parameter}
\end{table}

\paragraph{LLS Model}
The initialization of the stock return is performed by creating an artificial history of stock returns. The artificial history is modeled as a Gaussian random variable with mean
$\mu_h$ and standard deviation $\sigma_h$. Furthermore, we have to point out that the increments of the dividend is deterministic, if $z_1=z_2$ holds. We used the C++  standard random number generator for all simulations of the LLS model if not otherwise stated. 
\begin{table}
	\begin{subtable}[b]{0.45\textwidth}
		\begin{center}
			\begin{tabular}{|c||c|}
			\hline
			Parameter & Value\\
			\hline
			\hline
			$N$ & $100$\\
			\hline
			$m_i$& $15$\\
			\hline 
			$\sigma_{\gamma}$ & $ 0 $ or $0.2$\\
			\hline
			$r$& $0.04$ \\
			\hline 
			$z_1=z_2$& $0.05$\\
			\hline
			$\Delta t$ &$ 1$\\
			\hline
			time steps & 200 \\
			\hline
			\end{tabular}
		\end{center}
		\caption{Parameters of LLS model.}
	\end{subtable}
	\hspace{0.5cm}
	\begin{subtable}[b]{0.45\textwidth}
		\begin{center}
			\begin{tabular}{|c||c|}
			\hline
			 Variable & Initial Value\\
			\hline
			\hline
			$\mu_h$ & $0.0415$\\
			\hline
			$\sigma_h$ & $0.003$\\
			\hline
			$\gamma(t=0)$ & $0.4$\\
			\hline
			 $w_i(t=0)$ & $1000$ \\
			 \hline
			 $n_i (t=0)$ & $100$ \\
			 \hline
			 $S (t=0)$ & $4$ \\
			 \hline
			 $D (t=0)$ & $0.2$ \\
			 \hline
			\end{tabular}
		\end{center}
		\caption{Initial values of LLS model.}
	\end{subtable}
	\caption{Basic setting of the LLS model.} \label{LLS-basic}
\end{table}

\begin{table}
	\begin{subtable}[b]{0.45\textwidth}
		\begin{center}
			\begin{tabular}{|c||c|}
			\hline
			Parameter & Value\\
			\hline
			\hline
			$N$ & $99$\\
			\hline
			$m_i $&  $10,\ 1 \leqslant i \leqslant 33$  \\
			                  & $141,\ 34 \leqslant i \leqslant 66$ \\
			                  & $256,\ 67 \leqslant i \leqslant 99$ \\
			\hline 
			$\sigma_{\gamma}$ & $ 0.2 $\\
			\hline
			$r$& $0.0001$ \\
			\hline 
			$z_1=z_2$& $0.00015$\\
			\hline
			$\Delta t$ &$ 1 $\\
			\hline 
			time steps & $20,000$\\
			\hline
			\end{tabular}
		\end{center}
		\caption{Parameters of LLS model.} 
	\end{subtable}
	\hfill
	\begin{subtable}[b]{0.45\textwidth}
		\begin{center}
			\begin{tabular}{|c||c|}
			\hline
			 Variable & Initial Value\\
			\hline
			\hline
			  $\mu_h$ & $0.0415$ \\
			  \hline
			  $\sigma_h$ &$ 0.003$ \\
			  \hline
			  $\gamma_i(t=0)$ & $0.4$ \\
			  \hline
			  $w_i (t=0)$ & $1000$ \\
			  \hline
			   $n_i (t=0)$ & $100$ \\
			   \hline
			    $S (t=0)$ & $4 $\\
			    \hline
			   $D (t=0)$ & $0.004$ \\
			 \hline
			\end{tabular}
		\end{center}
		\caption{Initial values of LLS model.}
	\end{subtable}
	\caption{Setting for the LLS model (3 agent groups).}
	\label{LLS-3-agents}
\end{table}

\paragraph{Franke-Westerhoff Model}

\begin{table}[h!]
	\begin{subtable}[b]{0.45\textwidth}
    	\begin{center}
        	\begin{tabular}{|c||c|}
	            \hline
	            Parameter & Value\\
	            \hline
	            \hline
                $\phi$ & $1.0$\\
                \hline
                $\chi$ & $1.2$\\
                \hline
                $\eta$& $0.991$\\
                \hline
	            $\alpha_w$ & $1580$ \\
	            \hline
	            $\alpha_0$ & $0$ \\
	            \hline
	            $\alpha_h$ & $0$\\
	            \hline 
	            $\alpha_m$ & $0$\\
	            \hline
                $\sigma_f$ & $0.681$\\
	            \hline  
	            $\sigma_c$ & $1.724$\\
	            \hline
	            $\beta$ & $1$\\
	            \hline
	            $\nu$ & $-$\\
	            \hline
	            $P_f$ & $1$ \\
	            \hline
	            $\mu$ & $0.01$\\
	            \hline
	            time steps & $7,000$\\	            
	            \hline 
        	\end{tabular}
        \end{center}
        \caption{Parameters for DCA-W.}
	\end{subtable}
	\hfill
	\begin{subtable}[b]{0.45\textwidth}
		\begin{center}
	        \begin{tabular}{|c||c|}
	            \hline
	            Variable & Initial Value\\
	            \hline
	            \hline
	            $P(t=0)$ & 1 \\
	            \hline	            
	        \end{tabular}
	    \end{center}
	    \caption{Initial values DCA-W.}
	\end{subtable}
	\caption{DCA-W.}
    \label{dcaw-basic-parameter}
\end{table}

\begin{table}[h!]
	\begin{subtable}[b]{0.45\textwidth}
    	\begin{center}
        	\begin{tabular}{|c||c|}
	            \hline
	            Parameter & Value\\
	            \hline
	            \hline
                $\phi$ & $1.0$\\
                \hline
                $\chi$ & $0.9$\\
                \hline
                $\eta$& $0.987$\\
                \hline
	            $\alpha_w$ & $2668$ \\
	            \hline
	            $\alpha_0$ & $2.1$ \\
	            \hline
	            $\alpha_h$ & $0$ \\
	            \hline
	            $\alpha_m$ & $0$\\
	            \hline
                $\sigma_f$ & $0.752$\\
	            \hline  
	            $\sigma_c$ & $1.726$\\
	            \hline
	            $\beta$& $1$\\
	            \hline
                $\nu$ & $-$\\
	            \hline
	            $P_f$ & $1$ \\
	            \hline
	            $\mu$ & $0.01$\\
	            \hline
	            time steps & $7,000$\\	            
	            \hline 
        	\end{tabular}
        \end{center}
        \caption{Parameters for DCA-WP.}
	\end{subtable}
	\hfill
	\begin{subtable}[b]{0.45\textwidth}
		\begin{center}
	        \begin{tabular}{|c||c|}
	            \hline
	            Variable & Initial Value\\
	            \hline
	            \hline
	            $P(t=0)$ & 1 \\
	            \hline	            
	        \end{tabular}
	    \end{center}
	    \caption{Initial values DCA-WP.}
	\end{subtable}
	\caption{DCA-WP.}
    \label{dcawp-basic-parameter}
\end{table}

\begin{table}[h!]
	\begin{subtable}[b]{0.45\textwidth}
    	\begin{center}
        	\begin{tabular}{|c||c|}
	            \hline
	            Parameter & Value\\
	            \hline
	            \hline
                $\phi$ & $1.0$\\
                \hline
                $\chi$ & $0.9$\\
                \hline
                $\eta$& $0.987$\\
                \hline
	            $\alpha_w$ & $2668$ \\
	            \hline
	            $\alpha_0$ & $2.1$ \\
	            \hline
	            $\alpha_h$ & $1.28$ \\
	            \hline 
	            $\alpha_m$ & $0$\\
	            \hline
                $\sigma_f$ & $0.741$\\
	            \hline  
	            $\sigma_c$ & $2.087$\\
	            \hline
	            $\beta$ & $1$\\
	            \hline 
	            $\nu$ & $-$\\
	            \hline
	            $P_f$ & $1$ \\
	            \hline
	            $\mu$ & $0.01$\\
	            \hline
	            time steps & $7,000$\\
	            \hline 
        	\end{tabular}
        \end{center}
        \caption{Parameters for DCA-WHP.}
	\end{subtable}
	\hfill
	\begin{subtable}[b]{0.45\textwidth}
		\begin{center}
	        \begin{tabular}{|c||c|}
	            \hline
	            Variable & Initial Value\\
	            \hline
	            \hline
	            $P(t=0)$ & 1 \\
	            \hline	            
	        \end{tabular}
	    \end{center}
	    \caption{Initial values DCA-WHP.}
	\end{subtable}
	\caption{DCA-WHP.}
    \label{dcawhp-basic-parameter}
\end{table}

\begin{table}[h!]
	\begin{subtable}[b]{0.45\textwidth}
    	\begin{center}
        	\begin{tabular}{|c||c|}
	            \hline
	            Parameter & Value\\
	            \hline
	            \hline
                $\phi$ & $0.12$\\
                \hline
                $\chi$ & $1.5$\\
                \hline
                $\eta$& $0$\\
                \hline
	            $\alpha_w$ & $0$ \\
	            \hline
	            $\alpha_0$ & $-0.327$ \\
	            \hline
	            $\alpha_h$ & $1.79$ \\
	            \hline 
	            $\alpha_m$ & $18.43$\\
	            \hline
                $\sigma_f$ & $0.758$\\
	            \hline  
	            $\sigma_c$ & $2.087$\\
	            \hline                
	            $\beta$& $1$\\
	            \hline
	            $\nu$ & $-$\\
	            \hline
	            $P_f$ & $1$ \\
	            \hline
	            $\mu$ & $0.01$\\
	            \hline
	            time steps & $7,000$\\	            
	            \hline 
                random seed & $2661$ \\
	            \hline
        	\end{tabular}
        \end{center}
        \caption{Parameters for DCA-HPM.}
	\end{subtable}
	\hfill
	\begin{subtable}[b]{0.45\textwidth}
		\begin{center}
	        \begin{tabular}{|c||c|}
	            \hline
	            Variable & Initial Value\\
	            \hline
	            \hline
	            $P(t=0)$ & 1 \\
	            \hline	            
	        \end{tabular}
	    \end{center}
	    \caption{Initial values DCA-HPM.}
	\end{subtable}
	\caption{DCA-HPM.}
    \label{dcahpm-basic-parameter}
\end{table}

\begin{table}[h!]
	\begin{subtable}[b]{0.45\textwidth}
    	\begin{center}
        	\begin{tabular}{|c||c|}
	            \hline
	            Parameter & Value\\
	            \hline
	            \hline
                $\phi$ & $1.15$\\
                \hline
                $\chi$ & $0.81$\\
                \hline
                $\eta$& $0.987$\\
                \hline
	            $\alpha_w$ & $1041$ \\
	            \hline
	            $\alpha_0$ & $0$ \\
	            \hline
	            $\alpha_h$ & $0$ \\
	            \hline 
	            $\alpha_m$ & $0$\\
	            \hline
                $\sigma_f$ & $0.715$\\
	            \hline  
	            $\sigma_c$ & $1.528$\\
	            \hline
	            $\beta$ & $-$\\
	            \hline
                $\nu$ & $0.05$\\
                \hline 
	            $P_f$ & $1$ \\
	            \hline
	            $\mu$ & $0.01$\\
	            \hline
	            time steps & $7,000$\\	            
	            \hline 
        	\end{tabular}
        \end{center}
        \caption{Parameters for TPA-W.}
	\end{subtable}
	\hfill
	\begin{subtable}[b]{0.45\textwidth}
		\begin{center}
	        \begin{tabular}{|c||c|}
	            \hline
	            Variable & Initial Value\\
	            \hline
	            \hline
	            $P(t=0)$ & 1 \\
	            \hline	            
	        \end{tabular}
	    \end{center}
	    \caption{Initial values TPA-W.}
	\end{subtable}
	\caption{TPA-W.}
    \label{tpaw-basic-parameter}
\end{table}

\begin{table}[h!]
	\begin{subtable}[b]{0.45\textwidth}
    	\begin{center}
        	\begin{tabular}{|c||c|}
	            \hline
	            Parameter & Value\\
	            \hline
	            \hline
                $\phi$ & $1.0$\\
                \hline
                $\chi$ & $0.83$\\
                \hline
                $\eta$& $0.987$\\
                \hline
	            $\alpha_w$ & $2668$ \\
	            \hline
	            $\alpha_0$ & $0.376$ \\
	            \hline
	            $\alpha_h$ & $0$ \\
	            \hline 
	            $\alpha_m$ & $0$\\
	            \hline
                $\sigma_f$ & $0.736$\\
	            \hline  
	            $\sigma_c$ & $1.636$\\
	            \hline
	            $\beta$& $-$\\
	            \hline
	            	$\nu$ & $0.05$\\
	            \hline
	            $P_f$ & $1$ \\
	            \hline
	            $\mu$ & $0.01$\\
	            \hline
	            time steps & $7,000$\\	            
	            \hline 
        	\end{tabular}
        \end{center}
        \caption{Parameters for TPA-WP.}
	\end{subtable}
	\hfill
	\begin{subtable}[b]{0.45\textwidth}
		\begin{center}
	        \begin{tabular}{|c||c|}
	            \hline
	            Variable & Initial Value\\
	            \hline
	            \hline
	            $P(t=0)$ & 1 \\
	            \hline	            
	        \end{tabular}
	    \end{center}
	    \caption{Initial values TPA-WP.}
	\end{subtable}
	\caption{TPA-WP.}
    \label{tpawp-basic-parameter}
\end{table}

\begin{table}[h!]
	\begin{subtable}[b]{0.45\textwidth}
    	\begin{center}
        	\begin{tabular}{|c||c|}
	            \hline
	            Parameter & Value\\
	            \hline
	            \hline
                $\phi$ & $0.18$\\
                \hline
                $\chi$ & $2.3$\\
                \hline
                $\eta$& $0.987$\\
                \hline
	            $\alpha_w$ & $0$ \\
	            \hline
	            $\alpha_0$ & $-0.161$ \\
	            \hline
	            $\alpha_h$ & $1.3$ \\
	            \hline 
	            $\alpha_m$ & $12.5$\\
	            \hline
                $\sigma_f$ & $0.79$\\
	            \hline  
	            $\sigma_c$ & $1.9$\\
	            \hline
	            $\beta$& $-$\\
	            \hline 
	            $\nu$ & $0.05$ \\
	            \hline
	            $P_f$ & $1$ \\
	            \hline
	            $\mu$ & $0.01$\\
	            \hline
	            time steps & $7,000$\\	            
	            \hline
                random seed & $2661$\\
	            \hline 
        	\end{tabular}
        \end{center}
        \caption{Parameters for TPA-HPM.}
	\end{subtable}
	\hfill
	\begin{subtable}[b]{0.45\textwidth}
		\begin{center}
	        \begin{tabular}{|c||c|}
	            \hline
	            Variable & Initial Value\\
	            \hline
	            \hline
	            $P(t=0)$ & 1 \\
	            \hline	            
	        \end{tabular}
	    \end{center}
	    \caption{Initial values TPA-HPM.}
	\end{subtable}
	\caption{TPA-HPM.}
    \label{tpahpm-basic-parameter}
\end{table}